\documentclass[usenatbib,iop,numberedappendix]{aeb_emulateapj_2010}
\usepackage{epsfig}
\usepackage{xspace}

  \SetSymbolFont{symbols}{bold}{OMS}{cmsy}{b}{n}
  \DeclareSymbolFont{bmisymbols}{OML}{cmm}{b}{it}
  \DeclareMathSymbol{\balpha}{0}{bmisymbols}{"0B}
  \DeclareMathSymbol{\bbeta}{0}{bmisymbols}{"0C}
  \DeclareMathSymbol{\bgamma}{0}{bmisymbols}{"0D}
  \DeclareMathSymbol{\bdelta}{0}{bmisymbols}{"0E}
  \DeclareMathSymbol{\bepsilon}{0}{bmisymbols}{"0F}
  \DeclareMathSymbol{\bzeta}{0}{bmisymbols}{"10}
  \DeclareMathSymbol{\boldeta}{0}{bmisymbols}{"11}
  \DeclareMathSymbol{\btheta}{0}{bmisymbols}{"12}
  \DeclareMathSymbol{\biota}{0}{bmisymbols}{"13}
  \DeclareMathSymbol{\bkappa}{0}{bmisymbols}{"14}
  \DeclareMathSymbol{\blambda}{0}{bmisymbols}{"15}
  \DeclareMathSymbol{\bmu}{0}{bmisymbols}{"16}
  \DeclareMathSymbol{\bnu}{0}{bmisymbols}{"17}
  \DeclareMathSymbol{\bxi}{0}{bmisymbols}{"18}
  \DeclareMathSymbol{\bpi}{0}{bmisymbols}{"19}
  \DeclareMathSymbol{\brho}{0}{bmisymbols}{"1A}
  \DeclareMathSymbol{\bsigma}{0}{bmisymbols}{"1B}
  \DeclareMathSymbol{\btau}{0}{bmisymbols}{"1C}
  \DeclareMathSymbol{\bupsilon}{0}{bmisymbols}{"1D}
  \DeclareMathSymbol{\bphi}{0}{bmisymbols}{"1E}
  \DeclareMathSymbol{\bchi}{0}{bmisymbols}{"1F}
  \DeclareMathSymbol{\bpsi}{0}{bmisymbols}{"20}
  \DeclareMathSymbol{\bomega}{0}{bmisymbols}{"21}
  \DeclareMathSymbol{\bvarepsilon}{0}{bmisymbols}{"22}
  \DeclareMathSymbol{\bvartheta}{0}{bmisymbols}{"23}
  \DeclareMathSymbol{\bvarpi}{0}{bmisymbols}{"24}
  \DeclareMathSymbol{\bvarrho}{0}{bmisymbols}{"25}
  \DeclareMathSymbol{\bvarsigma}{0}{bmisymbols}{"26}
  \DeclareMathSymbol{\bvarphi}{0}{bmisymbols}{"27}

\newcommand{\elct}{\mathrm{e}}
\newcommand{\rmn}{\mathrm}
\newcommand{\dd}{\mathrm{d}}

\newcommand{\atsz}{A_{\mathrm{tSZ}}}

\newcommand{\cltsz}{C_{\ell,\mathrm{tSZ}}}

\slugcomment{Submitted to ApJ}

\shorttitle{Deconstructing the tSZ Power Spectrum}
\shortauthors{{Battaglia, Bond, Pfrommer, Sievers}}

\begin{document}

\title{On the Cluster Physics of Sunyaev-Zel'dovich Surveys II:\\
Deconstructing the Thermal SZ Power Spectrum}

\author{N. Battaglia\altaffilmark{1,2,3}, J. R. Bond\altaffilmark{2}, C. Pfrommer\altaffilmark{4,2},  J. L. Sievers\altaffilmark{2,5}}

\altaffiltext{1}{Department of Astronomy and Astrophysics, University of Toronto, 50 St George, Toronto ON, Canada, M5S 3H4}
\altaffiltext{2}{Canadian Institute for Theoretical Astrophysics, 60 St George, Toronto ON, Canada, M5S 3H8}
\altaffiltext{3}{McWilliams Center for Cosmology, Carnegie Mellon University,
Department of Physics, 5000 Forbes Ave., Pittsburgh PA, USA, 15213}
\altaffiltext{4}{Heidelberg Institute for Theoretical Studies, Schloss-Wolfsbrunnenweg 35, D-69118 Heidelberg, Germany}
\altaffiltext{5}{Joseph Henry Laboratories of Physics, Jadwin Hall, Princeton University, Princeton NJ, USA, 08544}

\begin{abstract}

Secondary anisotropies in the cosmic microwave background are a
treasure-trove of cosmological information.  Interpreting current
experiments probing them are limited by theoretical uncertainties
rather than by measurement errors.  Here we focus on the secondary
anisotropies resulting from the thermal Sunyaev-Zel'dovich (tSZ)
effect; the amplitude of which depends critically on the average
thermal pressure profile of galaxy groups and clusters.  To this end,
we use a suite of hydrodynamical TreePM-SPH simulations that include
radiative cooling, star formation, supernova feedback, and energetic
feedback from active galactic nuclei (AGN).  We examine in detail how
the pressure profile depends on cluster radius, mass, and redshift and
provide an empirical fitting function. We employ three different
approaches for calculating the tSZ power spectrum: an analytical
approach that uses our pressure profile fit, a semi-analytical method
of pasting our pressure fit onto simulated clusters, and a direct
numerical integration of our simulated volumes. We demonstrate that
the detailed structure of the intracluster medium and cosmic web
affect the tSZ power spectrum. In particular, the substructure and
asphericity of clusters increase the tSZ power spectrum by $10-20$\%
at $\ell \sim 2000-8000$, with most of the additional power being
contributed by substructures.  The contributions to the power spectrum
from radii larger than $R_{500}$ is $\sim 20$\% at $\ell = 3000$, thus
clusters interiors ($r < R_{500}$) dominate the power spectrum
amplitude at these angular scales.

\end{abstract}


\section{Introduction}

As cosmic microwave background (CMB) photons travel through the diffuse hot gas
comprising the bulk of baryons in galaxy clusters, a fraction of them are
upscattered by the gas in a process called the thermal Sunyaev-Zel'dovich (tSZ)
effect \citep{1970Ap&SS...7....3S}.  This scattering produces a unique spectral
signature in the CMB, with a decrement in thermodynamic temperature below $\nu
\sim 220$~GHz, and an excess above.  The tSZ effect is typically seen on
arc-minute scales, and is referred to as a secondary anisotropy, as it
originates between us and the surface of last scattering, unlike the primary CMB
anisotropies.  In the non-relativistic limit, the tSZ is directly proportional
to the integrated electron pressure along the line-of-sight.  It typically
traces out the spatial distribution of clusters and groups, since the hot
intracluster medium (ICM) dominates the line-of-sight pressure integral.  Thus,
the tSZ provides an excellent tool to examine the bulk of cluster baryons.
Found at the intersections of filaments in the cosmic web
\citep{1996Natur.380..603B}, clusters form at sites of constructive interference
of long waves in the primordial density fluctuations, the coherent peak-patches
\citep{1986ApJ...304...15B,1996ApJS..103....1B}. Clusters are sign posts for the
growth of structure in the Universe, and are a potentially powerful tool for
probing underlying cosmological parameters, such as $w$, the dark energy
pressure-to-density ratio.

The angular power spectrum of the tSZ effect is extremely sensitive to
cosmological parameters like $\sigma_8$, the root mean square (RMS) amplitude of
the (linearized) density fluctuations on 8$h^{-1}$ Mpc scales. In fact, the
amplitude of the tSZ power spectrum scales at least as steeply as the seventh
power of $\sigma_8$
\citep{2002ASPC..257...15B,2002MNRAS.336.1256K,2005ApJ...626...12B,2011ApJ...727...94T}
and improving the constraints on $\sigma_8$ will aid in breaking the
degeneracies found between $\sigma_8$ and $w$ when using only primary CMB
constraints. An advantage of using the tSZ angular power spectrum over counting
clusters is that no explicit measurement of cluster masses is required. Also,
lower mass, and therefore fainter, clusters that may not be significantly
detected as individual objects in CMB maps contribute to this statistical
signal.  However, disadvantages of using the tSZ angular power spectrum include
potential contamination from point sources and that no redshift information from
the clusters is used.

Previous observations by the Berkeley-Illinois-Maryland Association
\citep[BIMA,][]{2006ApJ...647...13D}, the Atacama Path-finding Experiment
\citep[APEX-SZ,][]{2009ApJ...701.1958R}, the Quest at DASI
\citep[QUaD,][]{2009ApJ...700L.187F}, Arc-minute Cosmology Bolometer Array
Receiver \citep[ACBAR,][]{2009ApJ...694.1200R}, and the Cosmic Background Imager
\citep[CBI,][]{2009arXiv0901.4540S} all measured excess power above that
expected from primary anisotropies, which have been attributed to some
combination of the tSZ effect and point source contamination. The measurements
from these experiments provided upper limits to the tSZ power spectrum
amplitude.  More recently, the Atacama Cosmology Telescope
\citep[ACT,][]{2010ApJ...722.1148F,2010arXiv1009.0866D} and the South Pole
Telescope
\citep[SPT,][]{2010ApJ...719.1045L,2010arXiv1012.4788S,2011arXiv1105.3182K} have
detected the SZ effect in the CMB power spectrum\footnote{The Planck
  collaboration has released some early SZ science
  \citep[e.g.,][]{2011arXiv1101.2026P,2011arXiv1101.2024P,2011arXiv1101.2043P},
  but to-date there have been no power spectrum results.}.  The results from ACT
and SPT emphasize that the ``sweet spot'' for measuring the tSZ signal is
between $\ell \sim 2000 - 4000$.  Silk damping \citep{1968ApJ...151..459S}
suppresses the power of primary anisotropies so that their contributions to the
power spectrum are much smaller than the tSZ contribution at even higher $\ell$.
At these scales there are important additional contributions to the power
spectrum from the kinetic SZ (kSZ) effect, which arises from motions of ionized
gas with respect to the CMB rest frame, as well as dusty star-forming galaxies
and the radio galaxies, both of which appear as point sources. All these signals
increase the uncertainty when determining the tSZ power spectrum, and hence the
parameters derived therefrom.

Three main tools have been used to estimate the tSZ power spectrum:
Analytic models, semi-analytical models, and numerical simulations.  They have been used
to derive several different templates for the predicted tSZ power
spectrum
\citep[e.g.,][]{1988MNRAS.233..637C,1993ApJ...405....1M,2000MNRAS.317...37D,2000PhRvD..61l3001R,2001ApJ...558..515H,2001ApJ...549...18Z,2001ApJ...549..681S,2002MNRAS.336.1256K,2002ApJ...577..555Z,2005ApJ...626...12B,2006MNRAS.370.1309S,2006MNRAS.370.1713S,2010ApJ...725...91B,2010ApJ...725.1452S,2011ApJ...727...94T,2011arXiv1106.3208E}.
There are both shape and amplitude differences between these three
approaches that compute the tSZ power spectrum; comparisons are
required to understand these differences
\citep{2000PhRvD..61l3001R}. At the base of these differences is the
cluster electron pressure profile, since it is a crucial and uncertain
component in the analytical thermal SZ power spectrum calculation. The
electron pressure profile is directly related to the total thermal
energy in a cluster and is sensitive to all the complicated
gastrophysics of the ICM. For example, some of the ICM processes that
should be included are radiative cooling, star-formation, energetic
feedback from AGN and massive stars, non-thermal pressure support,
magnetic fields, and cosmic rays.  Deviations from an average pressure
profile, i.e., cluster substructure and asphericity will also
contribute to the tSZ power spectrum. But how much?

The inclusion of AGN feedback is vital to any tSZ power spectrum template
\citep{2010ApJ...725...91B}. Furthermore, an energetic feedback source (AGN
feedback being the most popular) seems to be an important addition to any
hydrodynamical simulation, since simulations with {\em only} radiative cooling
and supernova feedback have problems with excessive over-cooling in cluster
centers \citep[e.g.,][]{2000ApJ...536..623L}. This over-cooling results in too
many stars being produced out of ICM gas reservoir, which alters the thermal and
hydrodynamic structure of ICM in a way that is inconsistent with observational
data.

In this paper we present a detailed comparison of the three approaches
used to calculated the thermal SZ angular power spectrum.  This
comparison allows us to identify and quantify the differences between
each method. Section \ref{sec:sims} briefly summarizes the
simulations used in this work and Section \ref{sec:tSZthry} outlines
the calculation of the analytical tSZ angular power spectrum. In
Sections \ref{sec:prof} and \ref{sec:PS} we present our results for
numerical average thermal pressure profiles and a detailed analysis of
the tSZ power spectrum, respectively. In Section \ref{sec:sig8} we
provide updated constraints on $\sigma_8$ using the new ACT and SPT
measurements of the CMB power spectrum at high $\ell$, and we
summarize our results and conclude in Section \ref{sec:con}.

\section{Cosmological simulations and cluster data set}
\label{sec:sims}

We use a modified version of the smoothed particle hydrodynamical
(SPH) code GADGET-2 \citep{2005MNRAS.364.1105S} to simulate
cosmological volumes.  We use a suite of 10 simulations with periodic
boundary conditions, box size $165\,h^{-1}\,\rmn{Mpc}$, and with equal
numbers of dark matter and gas particles $N_\rmn{DM}=N_{\mathrm{gas}}
= 256^3$.  We adopt a flat tilted $\Lambda$CDM cosmology, with total matter
density (in units of the critical)
$\Omega_{\mathrm{m}}=\Omega_{\mathrm{DM}}+\Omega_{\mathrm{b}} = 0.25$,
baryon density $\Omega_{\mathrm{b}}$ = 0.043, cosmological constant
$\Omega_{\Lambda}$ = 0.75, a present day Hubble constant of $H_0 = 100
h \mbox{ km s}^{-1} \mbox{ Mpc}^{-1}$, a scalar spectral index of
the primordial power-spectrum $n_{\rmn{s}}$ = 0.96 and $\sigma_8$ =
0.8.  The particle masses are then $m_{\mathrm{gas}}= 3.2\times 10^9\,
h^{-1}\,\mathrm{M}_{\sun}$ and $m_{\mathrm{DM}}= 1.54\times 10^{10}\,
h^{-1}\,\mathrm{M}_{\sun}$.  The minimum gravitational smoothing
length is $\varepsilon_\rmn{s}=20\, h^{-1}\,$kpc; our SPH densities are
computed with 32 neighbours.

We include sub-grid models for {\it AGN feedback} \citep{2010ApJ...725...91B},
radiative cooling, star formation, and supernova feedback
\citep{1996ApJS..105...19K,1996ApJ...461...20H,2003MNRAS.339..289S}.  The AGN
feedback prescription included in the simulations \citep[for more details
see][]{2010ApJ...725...91B,battinprep} allows for lower resolution and hence can
be applied to large-scale structure simulations.  It couples the black hole
accretion rate to the global star formation rate (SFR) of the cluster, as
suggested by \citet{2005ApJ...630..167T}. The thermal energy is injected into
the ICM such that it is proportional to the star-formation within a given
spherical region.  Throughout this work we will refer to these simulations as
{\it AGN feedback}.

We adopt the standard working definition of cluster radii $R_{\Delta}$ as the
radius at which the mean interior density equals $\Delta$ times the {\em
  critical density}, $\rho_\rmn{cr}(z)$ (e.g., for $\Delta =200$ or 500).  For
clarity the critical density is

\begin{equation}
\rho_\rmn{cr}(z) = \frac{3H_0^2} {8\pi G} \left[\Omega_{\rmn{m}}(1 + z)^3 + \Omega_{\Lambda} \right].\\
\end{equation}

\noindent Here we have assumed a flat universe
($\Omega_m+\Omega_\Lambda = 1$) and are only interested at times
after the matter-radiation equality, i.e., the radiation term with
$\Omega_{\rmn{r}}$ is negligible.  It is important to note that all masses and
distances quoted in this work are given relative to $h = 0.7$, since most
observations are reported with this value of $h$. 

\section{The Analytic Calculations of tSZ Angular Power Spectrum}
\label{sec:tSZthry}
The tSZ can be adequately modelled as a random
distributed Poisson process on the sky
\citep{1988MNRAS.233..637C}\footnote{
Note that we are not including the contributions from the
clustering of clusters, since this is sub-dominant on scales of $\ell
> 300$ \citep{1999ApJ...526L...1K}.}.
There are two components in this model that are required for a
statistical representation of the secondary anisotropies: (1) The
number density for objects of a given class; and
(2) the profile of the same object and class, centered on its
position. We focus on groups and clusters, since
they are the dominant source of tSZ anisotropies.
This approach is referred to as the halo formalism
\citep[e.g.,][]{1988MNRAS.233..637C}.

The non-relativistic tSZ signal is the line-of-sight integration of
the electron pressure,

\begin{equation} 
\frac{\Delta T}{T} = f(\nu) y = f(\nu) \frac{\sigma_{\rmn{T}}}{m_\elct c^2}\int P_{\elct}(l) \dd l \, , 
\label{eq:ycomp}
\end{equation}

\noindent where $f(\nu)$ is the spectral function for the tSZ
\citep{1970Ap&SS...7....3S}, $y$ is the Compton-$y$ parameter,
$\sigma_{\rmn{T}}$ is the Thompson cross-section, $m_\elct$ is the
electron mass and $P_{\elct}$ is electron pressure\footnote{Here we
have ignored the temperature of the CMB, $T_{\rmn{CMB}}$, since
$T_{\rmn{CMB}} \ll T_{\elct}$, hence
$n_{\elct}\,k_{\rmn{B}}\left(T_{\elct} - T_{\rmn{CMB}}\right) \simeq
n_{\elct}\,k_{\rmn{B}}\,T_{\elct} = P_{\elct}$.}. For a fully ionized
medium, the thermal pressure $P_{\rmn{th}} = P_{\elct} ({5X_{\rmn{H}} +
3}) / 2(X_{\rmn{H}} + 1) = 1.932 P_{\elct}$, where $X_{\rmn{H}} =
0.76$ is the primordial hydrogen mass fraction, and $P_{\rmn{th}}$ is
the thermal pressure. 

We adopt the successful analytical {\it ansatz} for halo number
density as a function of mass

\begin{equation}
\frac{\dd n(M,z)}{\dd M} = \frac{\bar{\rho}_m}{2\,M^2} \frac{R(M)}{3\sigma(M,z)^2} \frac{\dd \sigma(M,z)^2}{\dd R(M)}\, f(\sigma(M,z))
\end{equation}
\noindent where $\sigma(M,z)$ is the RMS variance of the linear
density field smoothed on the scale of $R(M)$, and $f(\sigma)$ is a
functional form determined from N-body simulations
\citep[e.g.,][]{2001MNRAS.321..372J,2006ApJ...646..881W,2008ApJ...688..709T}. In
this work we use the mass function from \citet{2008ApJ...688..709T}
for the analytic calculations. Note that the tSZ power spectrum is
only mildly sensitive to the particulars of the mass function
\citep{2002MNRAS.336.1256K}.

The tSZ angular power spectrum at a multipole moment $\ell$ is 

\begin{equation}
C_{\ell,\rmn{tSZ}} = f(\nu)^2 \int \frac{\dd V}{\dd z} \dd z  \int \frac{\dd n(M,z)}{\dd M} |\tilde{y}_{\ell}(M,z)|^2 \dd M,
\end{equation}

\noindent where $\tilde{y}_{\ell}(M,z)$ is the form factor, which is
proportional to the Fourier transform of the projected electron pressure
profile, $P_{\elct}$.  We do not include higher order relativistic corrections
to $f(\nu)$ \citep{2006NCimB.121..487N}.

The functional form of $\tilde{y}_{\ell}(M,z)$ can be determined
empirically in observations or simulations \citep[e.g.,
][]{2007ApJ...668....1N,2010A&A...517A..92A}, or can be determined
analytically \citep[e.g.,
][]{2001MNRAS.327.1353K,2005ApJ...634..964O}.  Following
\citet{2002MNRAS.336.1256K} we compute $\tilde{y}_{\ell}(M,z)$
assuming spherical symmetry and using Limber's approximation,

\begin{equation}\label{eq:ff}
\tilde{y}_\ell(M,z) = \frac{4\pi r_s}{\ell_s^2}\frac{\sigma_{\mathrm{T}}}{m_{\mathrm{e}}c^2}\int x^2 P_{\elct}(x) \frac{\sin(\ell x/\ell_s)}{\ell x/\ell_s}\dd x,
\end{equation}

\noindent where $x \equiv r/r_s$ is a dimensionless radius, $\ell_s \equiv
D_{\mathrm{A}}/r_s $ is the corresponding angular wave number, and
$D_{\mathrm{A}}$ is the angular diameter distance.  We follow
\citet{1997ApJ...490..493N} in our definition of the scale radius in a cluster
with concentration $c_{\rm{NFW}}$, $r_{s} \equiv r_\rmn{vir}/c_{\rmn{NFW}}$.
Here we use a fitting formula for $c_{\rm{NFW}}$
from \citet{2008MNRAS.390L..64D} and the definition for the virial radius from
\citet{1998ApJ...495...80B},

\begin{equation}
r_\rmn{vir} = \left(\frac{3\,M_\rmn{vir}}{4\pi\,\Delta_{\rmn{cr}}(z)\,\rho_{\rmn{cr}}(z)}\right)^{1/3},
\end{equation}
\noindent where $\Delta_{\rmn{cr}}(z) = 18\pi^2 + 82[\Omega (z)- 1] -
39[\Omega (z)- 1]^2$ and $\Omega (z) =
\Omega_{\rmn{m}}(1+z)^3\,\left[\Omega_{\rmn{m}}(1+z)^3 +
\Omega_{\Lambda}\right]^{-1}$.

The dominant source of uncertainty in $C_{\ell,\rmn{tSZ}}$ comes from
$\tilde{y}_{\ell}(M,z)$, since one can easily calculate the volume
element for a given cosmology, and the mass function is known to
$5-10$\% \citep{2008ApJ...688..709T}. Thus, the pressure profile is
the critical input into the analytical tSZ angular power spectrum.  We
would ideally like to know $\tilde{y}_{\ell}(M,z)$ as well as we know
the mass function.  This requires an understanding of the detailed
physical processes which affect cluster pressure profiles.

The Gaussian and non-Gaussian variances of the power spectrum are also
calculated using the halo formalism
\citep{1996clss.conf..469B,2001PhRvD..64f3514C,2002MNRAS.336.1256K,2007ApJ...671...14Z,2009ApJ...702..368S},
again neglecting the clustering of clusters term. The full-sky variance is

\begin{equation}
\label{eq:clerr}
\sigma^2_{\ell\ell',\rmn{tSZ}} = \left[\frac{2(C_{\ell,\rmn{tSZ}})^2}{2\ell + 1}\delta_{\ell\ell'} + \frac{T_{\ell\ell'}}{4\pi}\right]
\end{equation}

\noindent where $T_{\ell\ell'}$ is the trispectrum; see
Equation~(\ref{eq:trispec}).  The variance is inversely proportional
to the sky area covered, so for a fraction $f_{\rmn{sky}}$ of the sky
covered, $\sigma^2_{\ell\ell',\rmn{tSZ}} \propto 1/f_{\rmn{sky}}$. In
this work we will present the diagonal part of the covariance; the
diagonal of the trispectrum is

\begin{equation}
\label{eq:trispec}
T_{\ell\ell,\rmn{tSZ}} = f(\nu)^4 \int \frac{\dd V}{\dd z} \dd z  \int \frac{\dd n(M,z)}{\dd M} |\tilde{y}_{\ell}(M,z)|^4 \dd M.
\end{equation}

\section{The Thermal Pressure Profile}
\label{sec:prof}

\begin{figure*}[thbp]
  \resizebox{0.5\hsize}{!}{\includegraphics{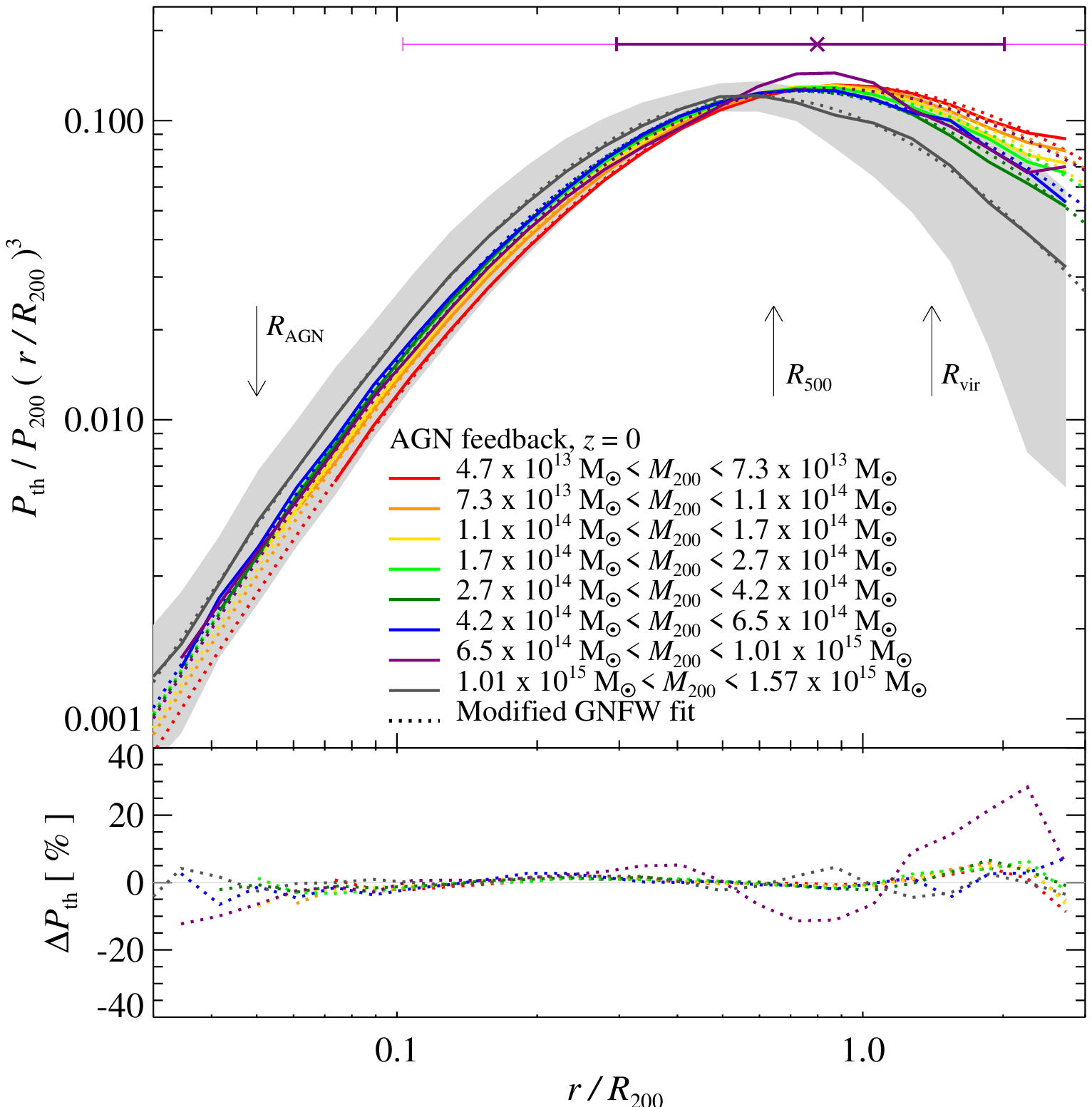}}%
  \resizebox{0.5\hsize}{!}{\includegraphics{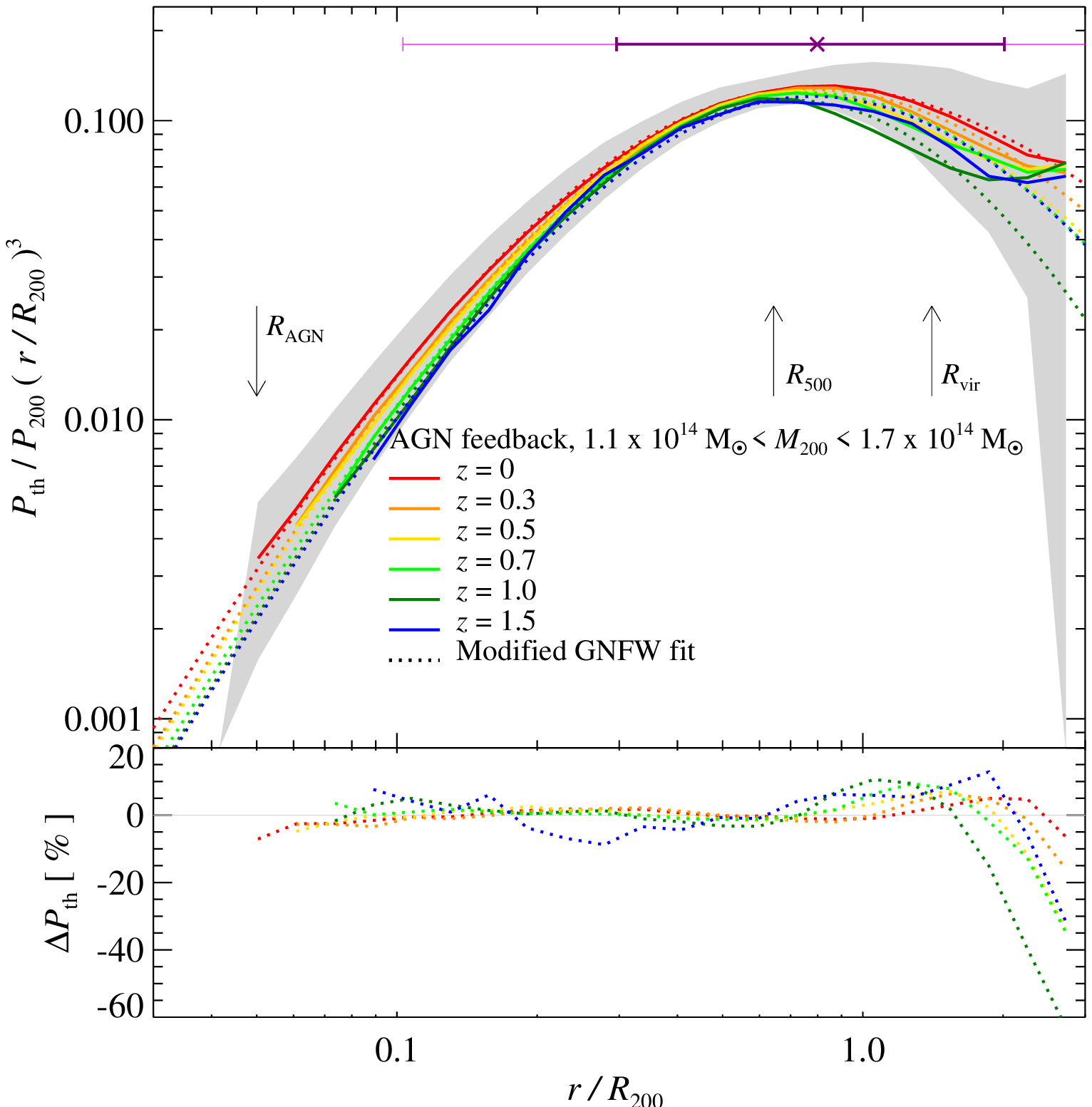}}\\
\caption[The individual fits to the thermal pressure profiles as
 functions of mass, redshift, and radius]{The normalized average
 pressure profiles and {\em parametrized fits} to these profiles from
 simulations with AGN feedback scaled by $(r/R_{200})^3$, in mass bins
 (left panel) and redshift bins (right panel). Here we have
 independently fit each mass and redshift bin. The grey band shows
 the standard deviation of the average cluster in the most massive bin
 (left) and lowest redshift bin (right).  In both panels we
 illustrate the radii that contribute 68\% and 95\% of the total thermal
 energy, $Y$, centered on the median, by horizontal purple and pink
 error bars. The bottom panels show the percent difference between the
 fits and the  average profiles. The generalized NFW profile with
 fixed $\alpha$ and $\gamma$ fits the average profiles well in the
 majority of the mass and redshift bins, with deviations within
 $\sim 5$\% of the mean. The upturns at large radii are due to
 contributions from nearby clusters and substructure.}
\label{fig:pth_mass_z}
\end{figure*}

\begin{figure*}[thbp]
  \resizebox{0.5\hsize}{!}{\includegraphics{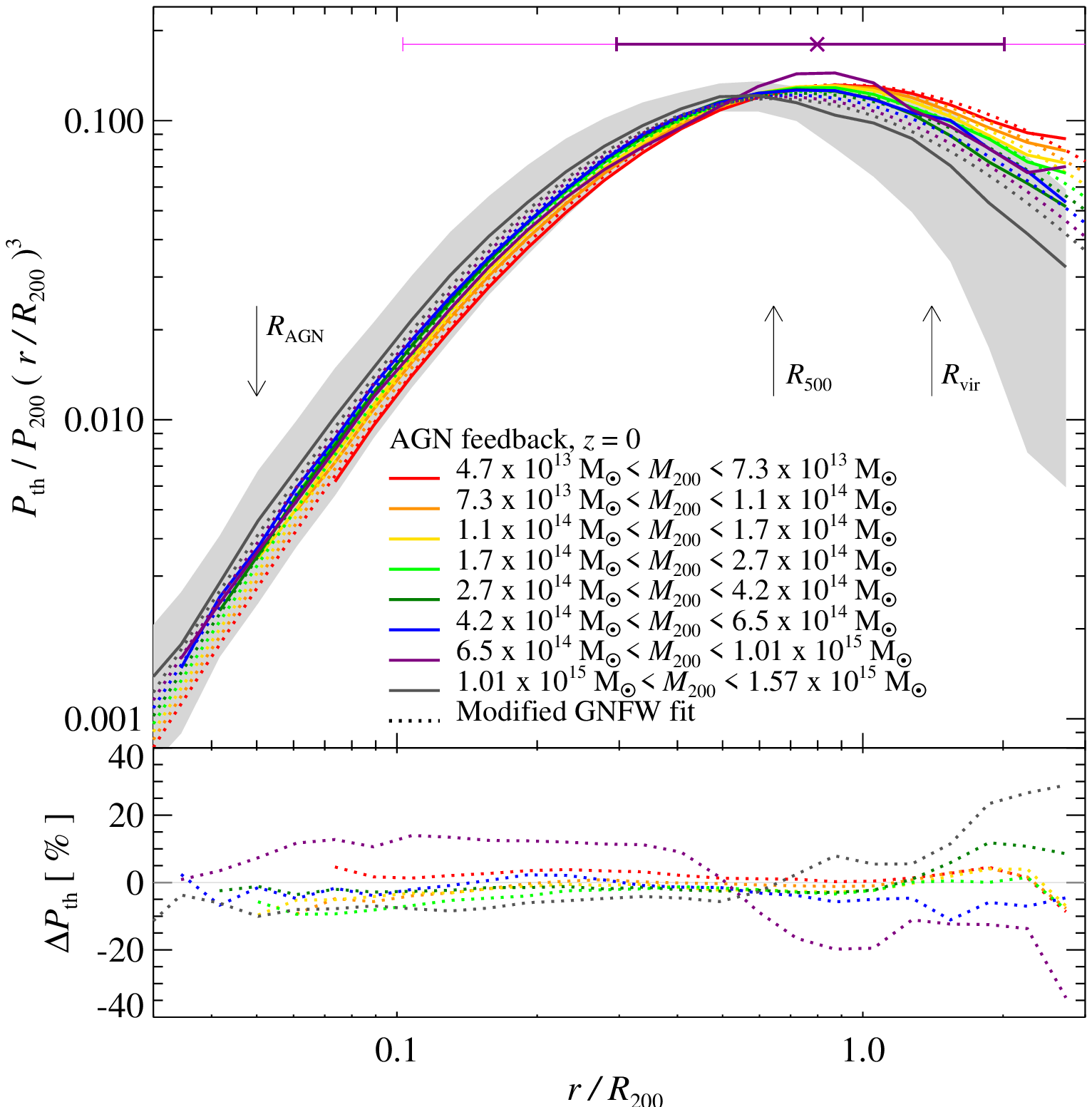}}%
  \resizebox{0.5\hsize}{!}{\includegraphics{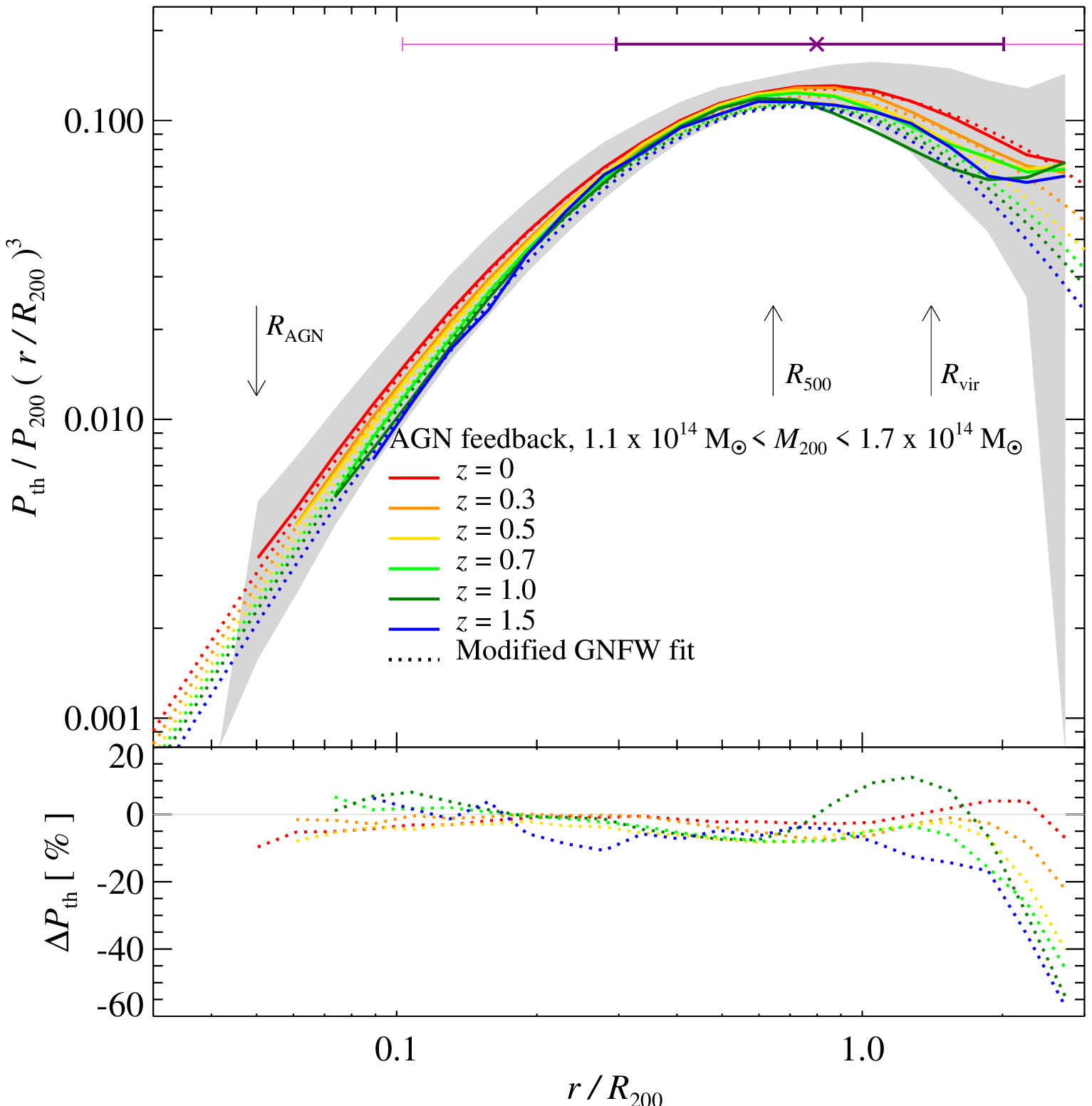}}\\
\caption[The constrained fits to the thermal pressure profiles as
  functions of mass, redshift, and radius]{The normalized average
  pressure profiles and {\em constrained fits} to these profiles from
  simulations with AGN feedback scaled by $(r/R_{200})^3$, for mass
  bins (left panel) and redshift bins (right panel).  The constrained
  fit is a global pressure profile, as described in Section \ref{sec:pfit2}, with
  parameters in Table \ref{tab:mzfit}.  It differs from the fits in
  Figure~\ref{fig:pth_mass_z}, where each bin was fit independently.
  The grey band shows the standard deviation of the average cluster in
  the most massive bin (left) and lowest redshift bin (right).  In
  both panels we illustrate the radii that contribute 68\% and 95\% of
  the total thermal energy, $Y$, centered on the median, by horizontal
  purple and pink error bars. The bottom panels show the percent
  differences between the constrained global fits and the average
  profiles.  The constrained fits match the average
  profiles well in the majority of the mass and redshift bins and the
  deviations are within $\sim 10$\% of the mean.  The upturns at large
  radii are due to contributions from substructure and nearby clusters.}
\label{fig:pth_mass_z_fit}
\end{figure*}

The cluster thermal pressure profile is the most uncertain component
of the tSZ power spectrum.  In this section we use a large sample of
clusters from hydrodynamical simulations and explore the mean cluster
profile and the subtle differences from self-similar scaling
\citep[e.g.,][]{1986MNRAS.222..323K,2005RvMP...77..207V}.  Comparisons
between the latest pressure profiles from analytics, observations, and
simulations have shown that they are in reasonable agreement with one
another
\citep{2010A&A...517A..92A,2010ApJ...725.1452S,2011ApJ...727...94T,2011ApJ...727L..49S}.
Previous work has shown that AGN feedback can alter the pressure
profiles, though the profiles are comparable to previous simulations
and observations \citep{2010ApJ...725...91B}.
We show the dependence of the
pressure profile on the cluster mass and redshift and explore
deviations from the self-similar scaling.

\subsection{Fitting Pressure Profiles from the Simulations}
\label{sec:pfit}

We apply the following four-step algorithm to compute the average
thermal pressure profiles in our simulations. First, we find all
clusters in a given snapshot using a friends-of-friends (FOF)
algorithm \citep{1982ApJ...257..423H} using a linking length of
0.2 and an $M_{FOF}$ mass cut of $1.4 \times 10^{13}M_{\sun}$. Second, starting with a 
position and radius derived from the FOF results, we find the final
cluster positions by recursively shrinking the radius of the sphere
examined, and re-center on its center of mass.  Given the cluster
center, we then calculate the spherical-overdensity mass and radius,
$M_{\Delta}$ and $R_{\Delta}$.
Third, we calculate the thermal pressure profile for the entire sample
of clusters in spherical shells, with the shells defined relative to
$R_{\Delta}$ (for the pressure profiles, we use $\Delta=200$).  To
facilitate profile comparisons and cluster stacking, we normalize each
profile by the self-similar amplitude for pressure $P_{\Delta} \equiv
G M_{\Delta} \Delta\, \rho_\rmn{cr}(z) f_{\mathrm{b}}/(2 R_{\Delta})$
\citep{1986MNRAS.222..323K,2005RvMP...77..207V}, with $f_{\mathrm{b}}
= \Omega_{\mathrm{b}} / \Omega_{\mathrm{m}}$.  Finally, we form a
weighted average of these profiles by stacking clusters in a given
redshift and mass bins. We use the integrated Compton-$y$ parameter as
our weighting function,

\begin{equation} 
Y_{\Delta} = \frac{\sigma_{\rmn{T}}}{m_\elct
c^2}\int^{R_{\Delta}}_0 P_{\elct}(r) 4\pi r^2\, \dd r \,  \propto E_{\rmn{th}} (< R_{\Delta})\, , 
\label{eq:Ydelta}
\end{equation}

\noindent The stacked average profiles $\bar{P}_{\rmn{th}}=
\left<P_{\rmn{th}}/P_{\Delta}\right>$ are then fit to a restricted
version of the generalized NFW profile,

\begin{equation}\label{eq:pfit} 
\bar{P}_{\rmn{fit}} = P_0
   \left(x/x_{\rmn{c}}\right)^{\gamma}\left[1 +
   \left(x/x_{\rmn{c}}\right)^{\alpha} \right]^{-\beta}, \ x \equiv
   r/R_{\Delta},
\end{equation}

\noindent where the fit parameters are a core-scale $x_{\rmn{c}}$, an amplitude
$P_0$ and a power law index $\beta$ for the asymptotic fall off of the profile.
There is substantial degeneracy between fit parameters, so we fix $\alpha = 1.0$
and $\gamma = -0.3$ \citep[as suggested
by][]{2007ApJ...668....1N,2010A&A...517A..92A}.  We find that fitting for all
parameters did not provide a significantly better fit than when $\alpha$ and
$\gamma$ were fixed. However, without fixing $\alpha$ and $\gamma$, a direct
comparison of fit parameters between different mass and redshift slices was not
meaningful.  We find the best-fit parameters using a non-linear least squares
Levenberg-Marquardt approach \citep{Levenberg1944Method,Marquardt_1963}.  We
weight each radial bin by the internal variance of the cluster profiles within
that bin.

In Figure \ref{fig:pth_mass_z}, we show the mass and redshift
dependence of the average cluster thermal pressure profile and the
corresponding parametrized fits to these profiles. We scale the
pressure profiles by $x^3$, such that the height corresponds to the
contribution per logarithmic radial interval to the total thermal energy
content of the cluster (cf. horizontal purple and pink error bars for
the radii that contribute 68\% and 95\% of the cluster thermal energy).
In the bottom panels of Figure \ref{fig:pth_mass_z}, we highlight the
residuals from the smoothed fitting function by showing the relative
difference in per cent, $\Delta P_{\rmn{th}} = 100
\left(P_{\rmn{fit}} - \bar{P}_{\rmn{th}}\right)/
\bar{P}_{\rmn{th}}$. The fitting function, Equation~\ref{eq:pfit},
provides an accurate fit over all mass and redshift ranges, with a
majority of the deviations from the average profile being $<5$\%.

We find that there are subtle dependencies on the cluster mass and redshift
(cf. Table \ref{tab:mzfit}), which suggests that neglecting these dependencies
would not yield the required $5-10$\% precision needed for calculations of tSZ
power spectrum.  We also find that there are contributions to the average
pressure profile at larger radii from substructure and nearby clusters, which
cause relative deviations from the mean profile  $>5$\%.  In a companion
paper, we also show that substructure affects the kinetic support in cluster
outskirts and the shape of the ICM shape at similar radii \citep{battinprep}.
In these regions (redshift dependent, but typically $ \gtrsim 2R_{200}$)
$P_{\rmn{fit}}$ often deviates from $\bar{P}_{\rmn{th}}$ by more than $5$\%. We
chose not to model this behavior because of two reasons. First, the problem of
double-counting SZ flux: the large volume contained within the radius that
contains 95\% of the total SZ flux, $r<4R_{200}$, necessarily leads to
overlapping volumes of neighboring clusters, especially at
high-redshift. Second, the total SZ flux of an increasing pressure profile,
scaled by $x^3$, does not converge and an arbitrarily chosen radial cutoff would
substantially contribute to the resulting power of the tSZ power spectrum.
Because we weight by the variance within radial bins, these contaminated regions
are naturally down-weighted in the profile fits.

\subsection{Constrained Thermal Pressure Profile Fits}
\label{sec:pfit2}

In this section we derive a global fit to our pressure profiles as a
function of mass and redshift.  We find treating each parameter as a
separable function of mass and redshift gives good results, with the
fit parameters constrained to be of the following form:  For generic
parameter $A$, we have
\begin{equation}\label{eq:pth_mfit}
  A = A_0\left(\frac{M_{200}}{10^{14}
    \,\mathrm{M}_{\sun}}\right)^{\alpha_{\mathrm{m}}} \left(1 + z\right)^{\alpha_{\mathrm{z}}}.
\end{equation}

\noindent For each of $P_0$, $\beta$, and $x_{\rm{c}}$, we find
$\alpha_{\mathrm{m}}$ by fitting to the $z=0$ snapshot, and we find
$\alpha_{\mathrm{z}}$ by fitting to clusters with $1.1\times10^{14} M_{\sun} <
M_{200} < 1.7\times10^{14} M_{\sun}$.  The weights used in the fits were the
inverse variance of the fit parameters when fitting each individual cluster in
that mass/redshift bin.  With these fit parameters, presented in Table
\ref{tab:mzfit}, and using Equations (\ref{eq:pfit}) and (\ref{eq:pth_mfit}), we
now have a global model for the average electron pressure as a function of
cluster radius, redshift, and mass.  Hereafter we refer to this global empirical
description as the constrained pressure profile. In Figure
\ref{fig:pth_mass_z_fit} we compare the constrained fits to the stacked
averages.  With fewer degrees of freedom, the constrained fits will naturally
not be as accurate as fitting each mass/redshift bin completely independently,
but we find that the mean recovered profile is accurate to $10$\% and
corresponds well to the accuracy with which we intend to measure the tSZ power
spectrum.

\begin{table}
  \caption[Mass and redshift dependencies of the average thermal
  pressure profile]{Mass and Redshift Fit Parameters from
  Eqns.~(\ref{eq:pfit})  and (\ref{eq:pth_mfit}).}
  \label{tab:mzfit}
  \begin{center}
   \leavevmode
   \begin{tabular}{lccc} 
     \hline \hline              
     Parameter & $A_{\mathrm{m}} = A_{\mathrm{z}}$ & $\alpha_{\mathrm{m}}$ & $\alpha_{\mathrm{z}}$ \\
     \hline
     $P_0$   & 18.1 & 0.154 &-0.758\\ 
     $x_{\rmn{c}}$   & 0.497 & -0.00865 & 0.731\\        
     $\beta$ & 4.35 & 0.0393 & 0.415\\ 
     \hline
    \end{tabular}
    \begin{quote}
      \noindent 
The input weights are chosen to be the inverse variances of fit
parameter values from the individual pressure fits for each cluster
within the bin.
    \end{quote}
  \end{center}
\end{table}

\begin{figure*}
  \begin{center}
    \resizebox{0.3333\hsize}{!}{\includegraphics{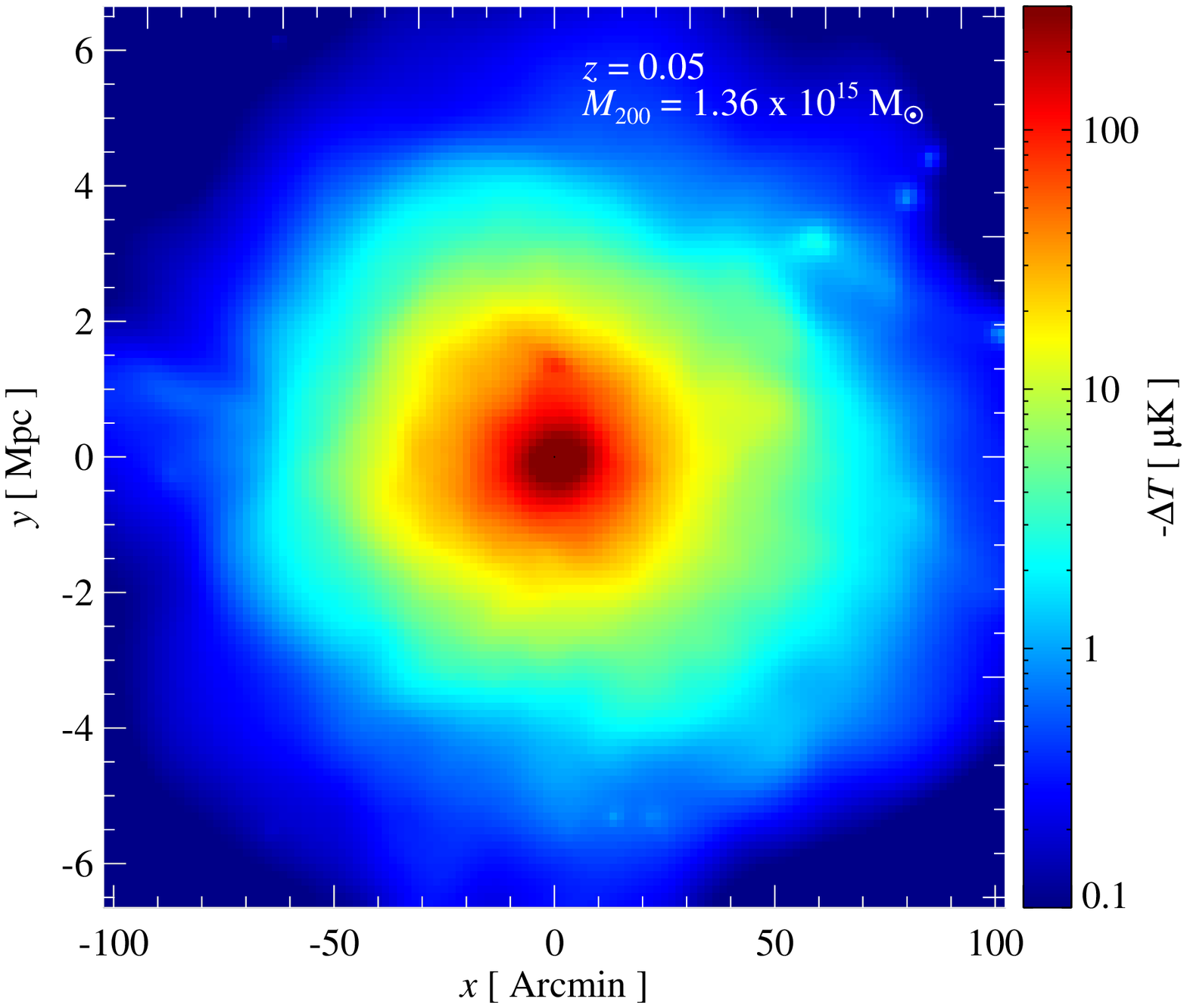}}%
    \resizebox{0.3333\hsize}{!}{\includegraphics{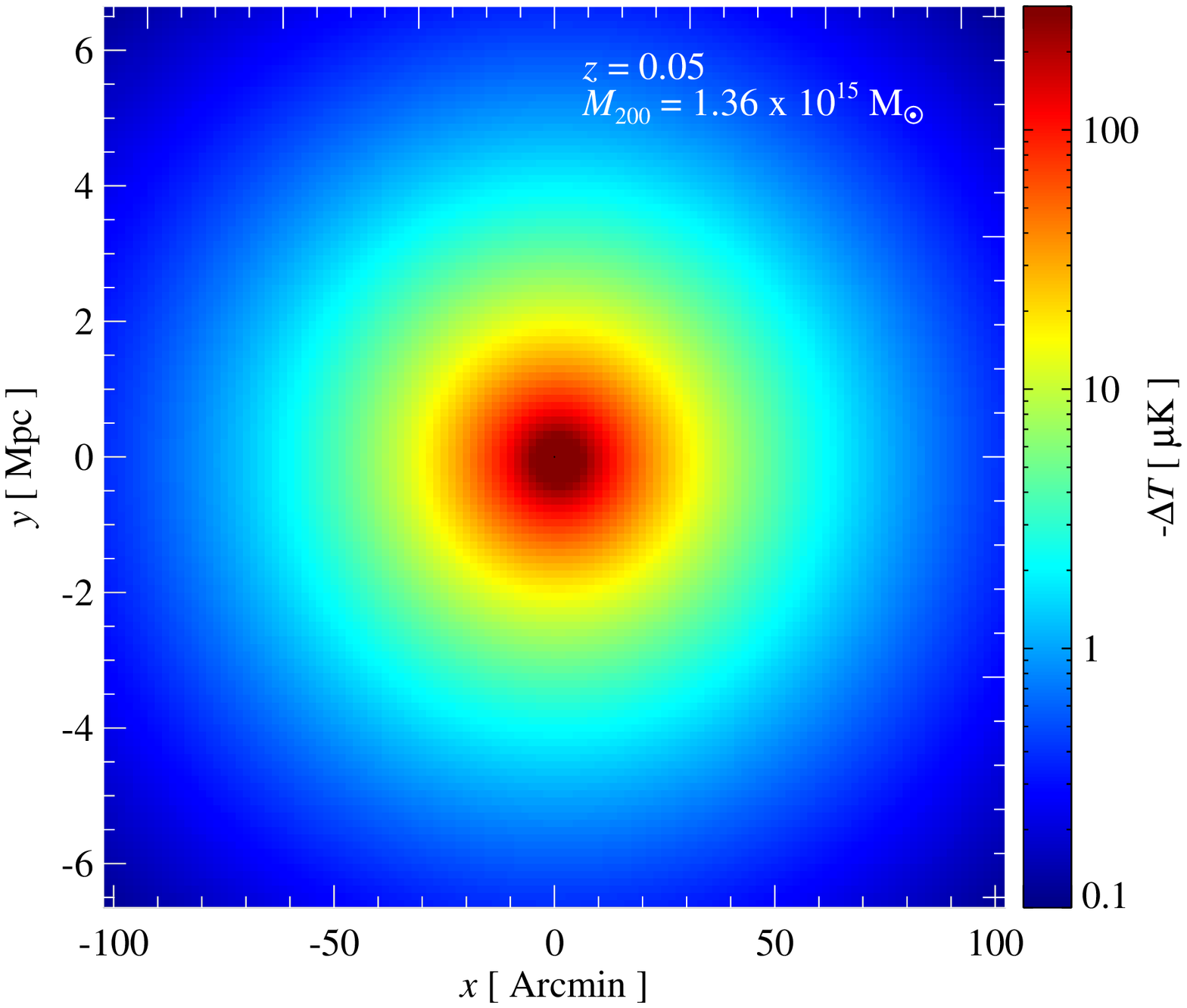}}%
    \resizebox{0.3333\hsize}{!}{\includegraphics{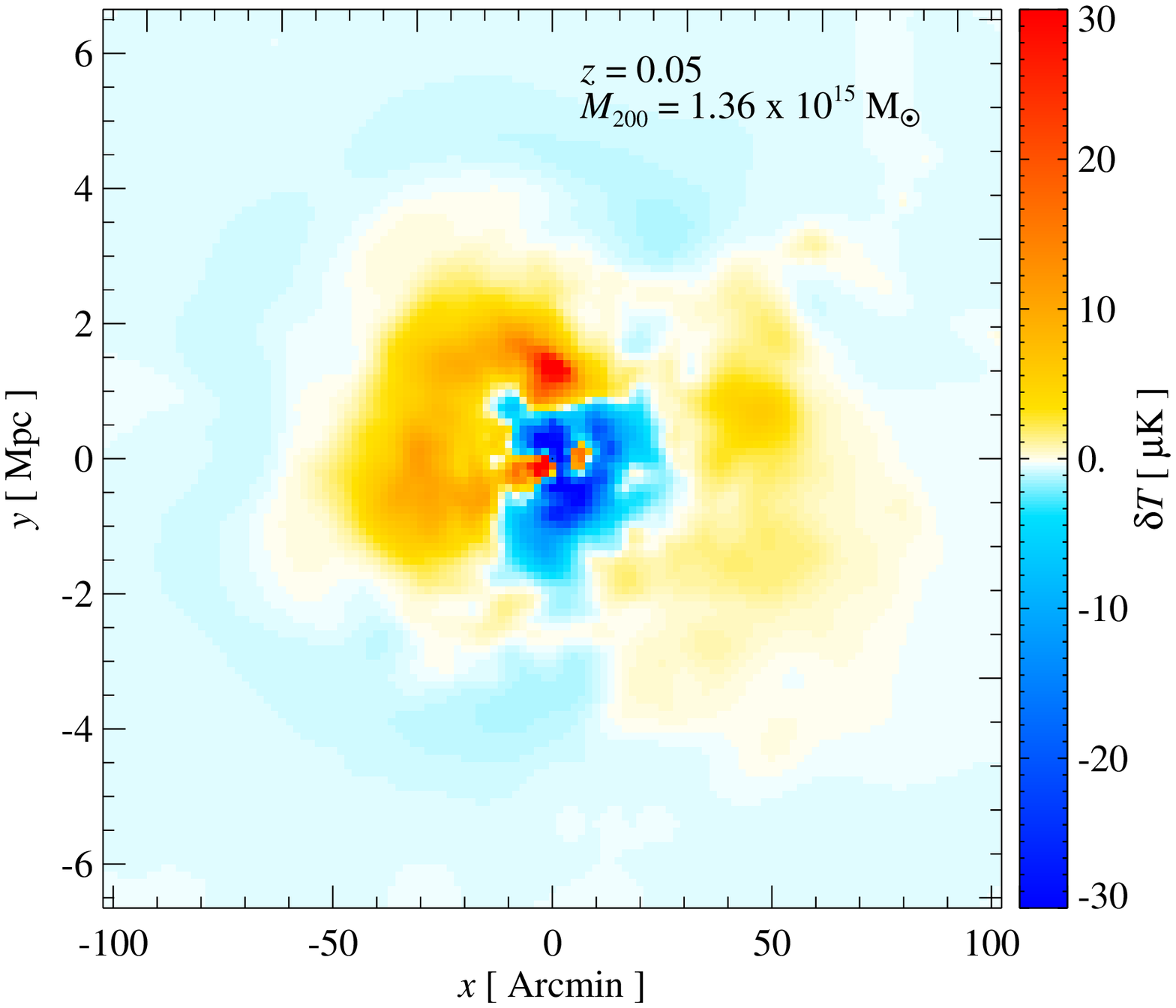}}\\
    \resizebox{0.3333\hsize}{!}{\includegraphics{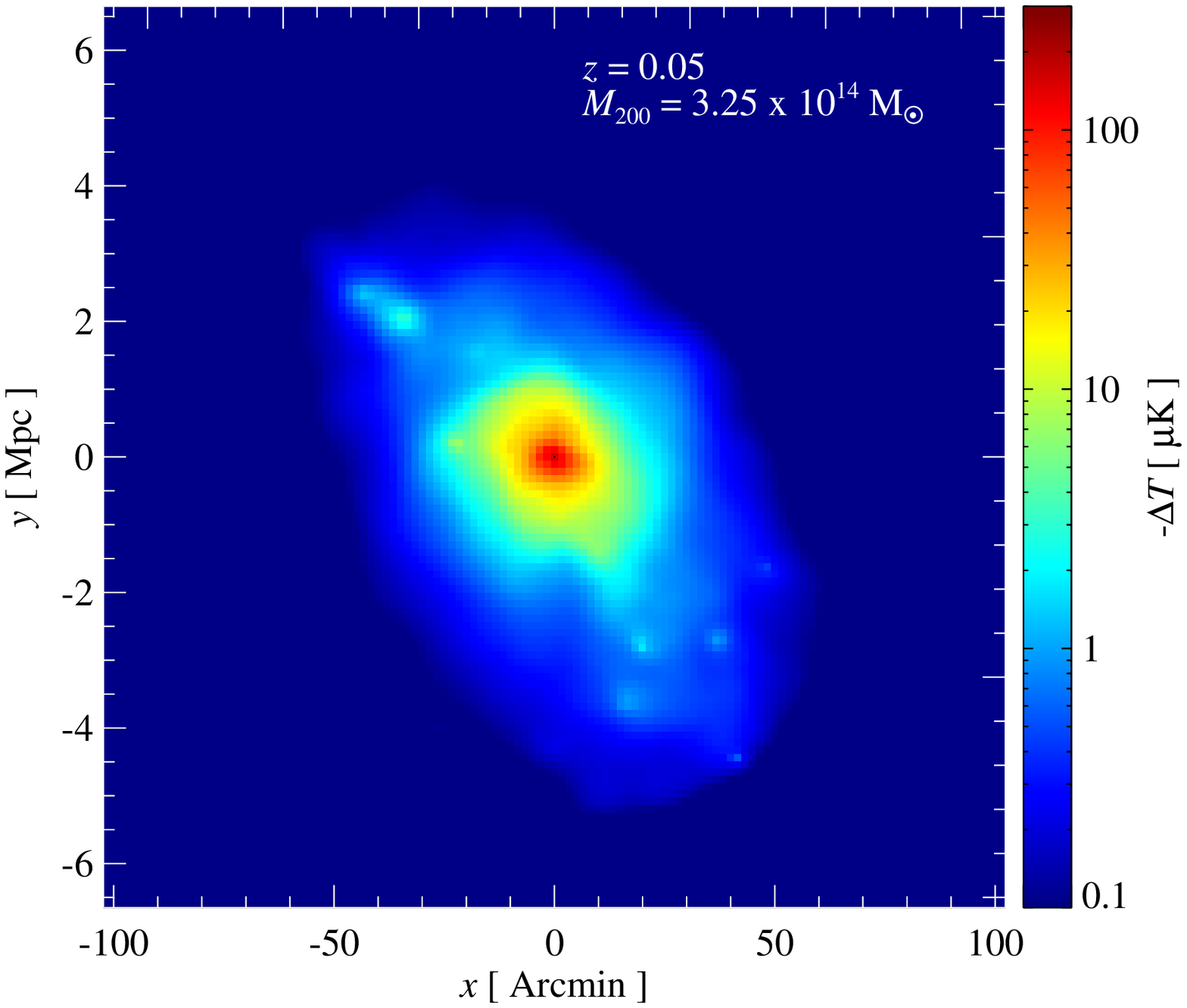}}%
    \resizebox{0.3333\hsize}{!}{\includegraphics{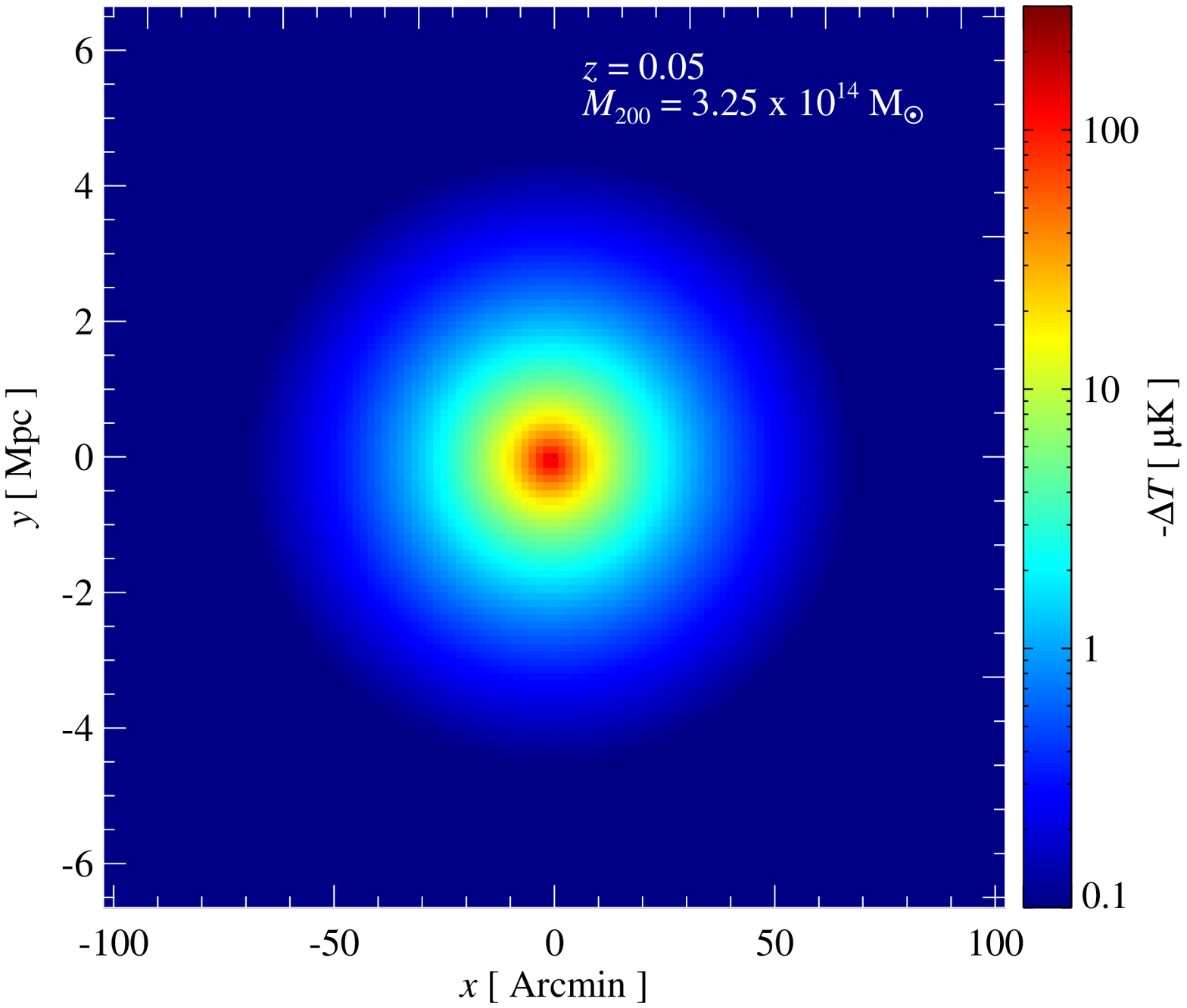}}%
    \resizebox{0.3333\hsize}{!}{\includegraphics{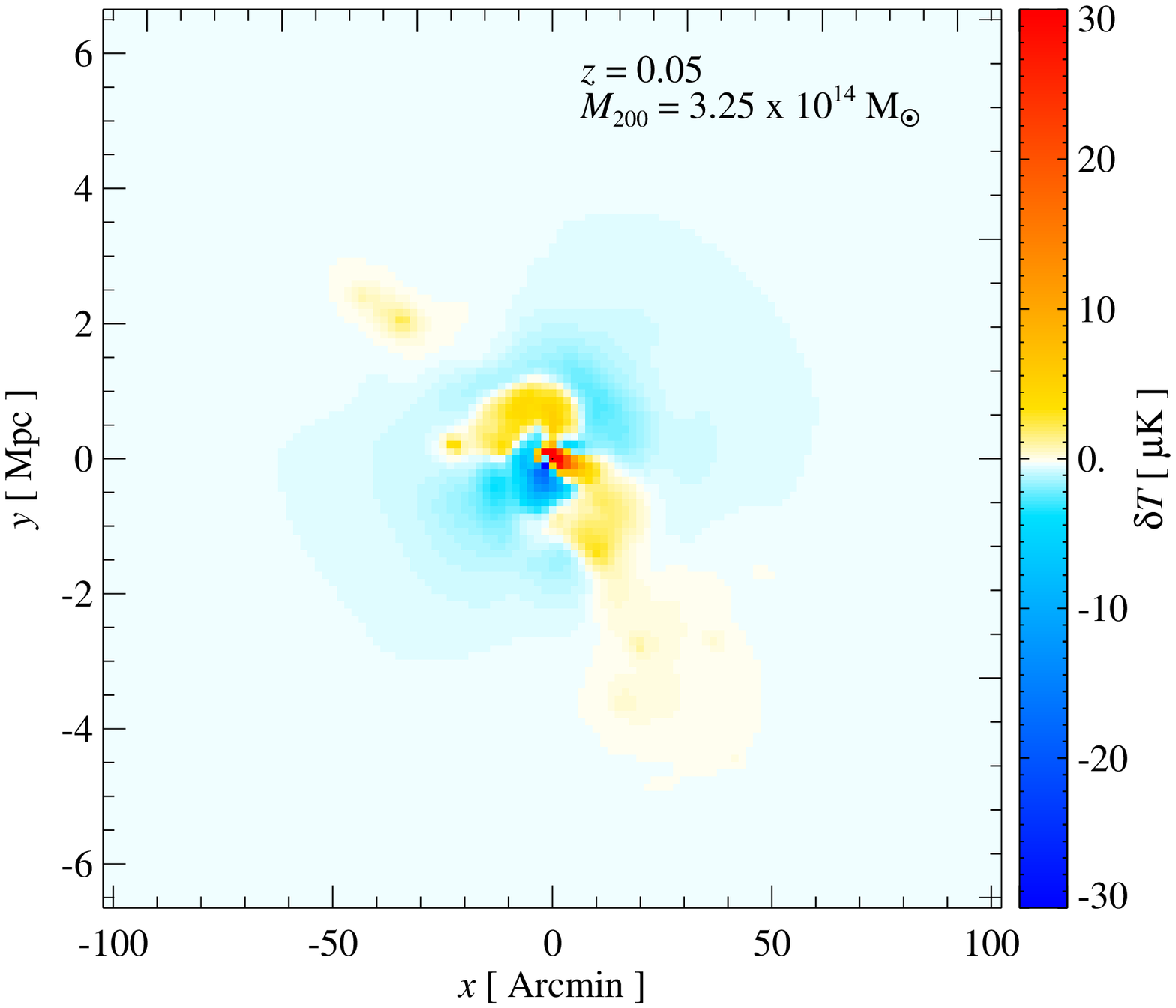}}\\
    \resizebox{0.3333\hsize}{!}{\includegraphics{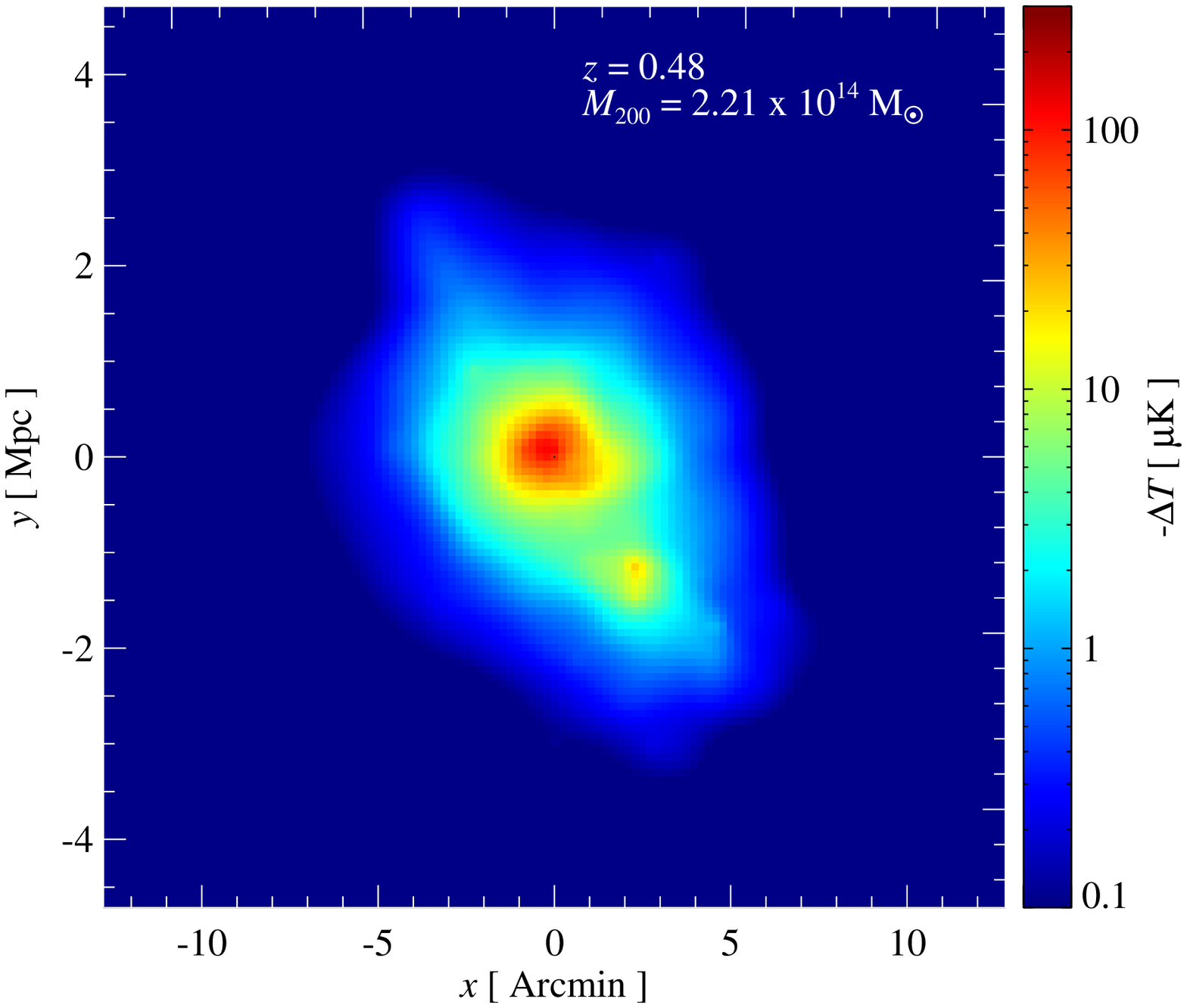}}%
    \resizebox{0.3333\hsize}{!}{\includegraphics{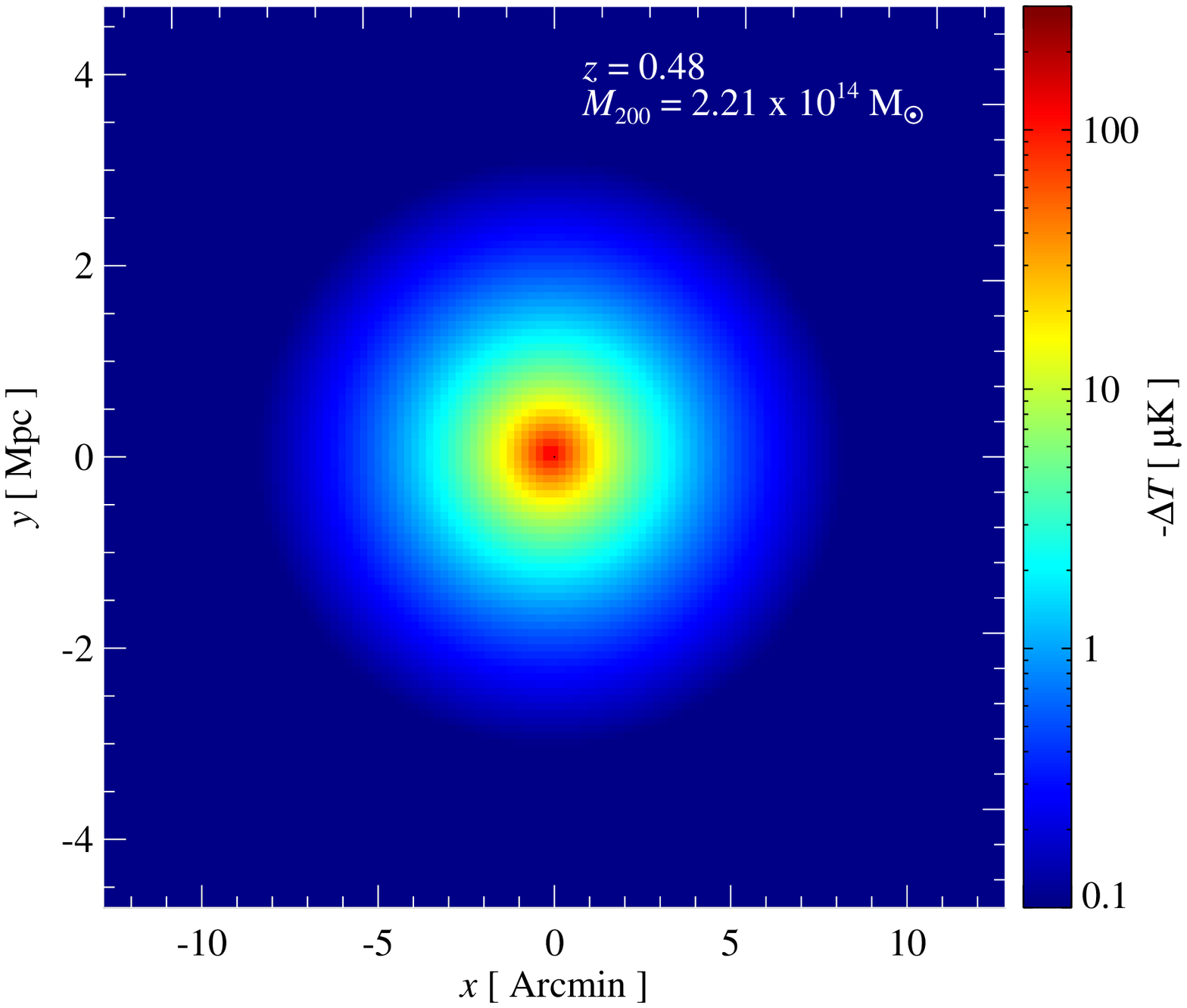}}%
    \resizebox{0.3333\hsize}{!}{\includegraphics{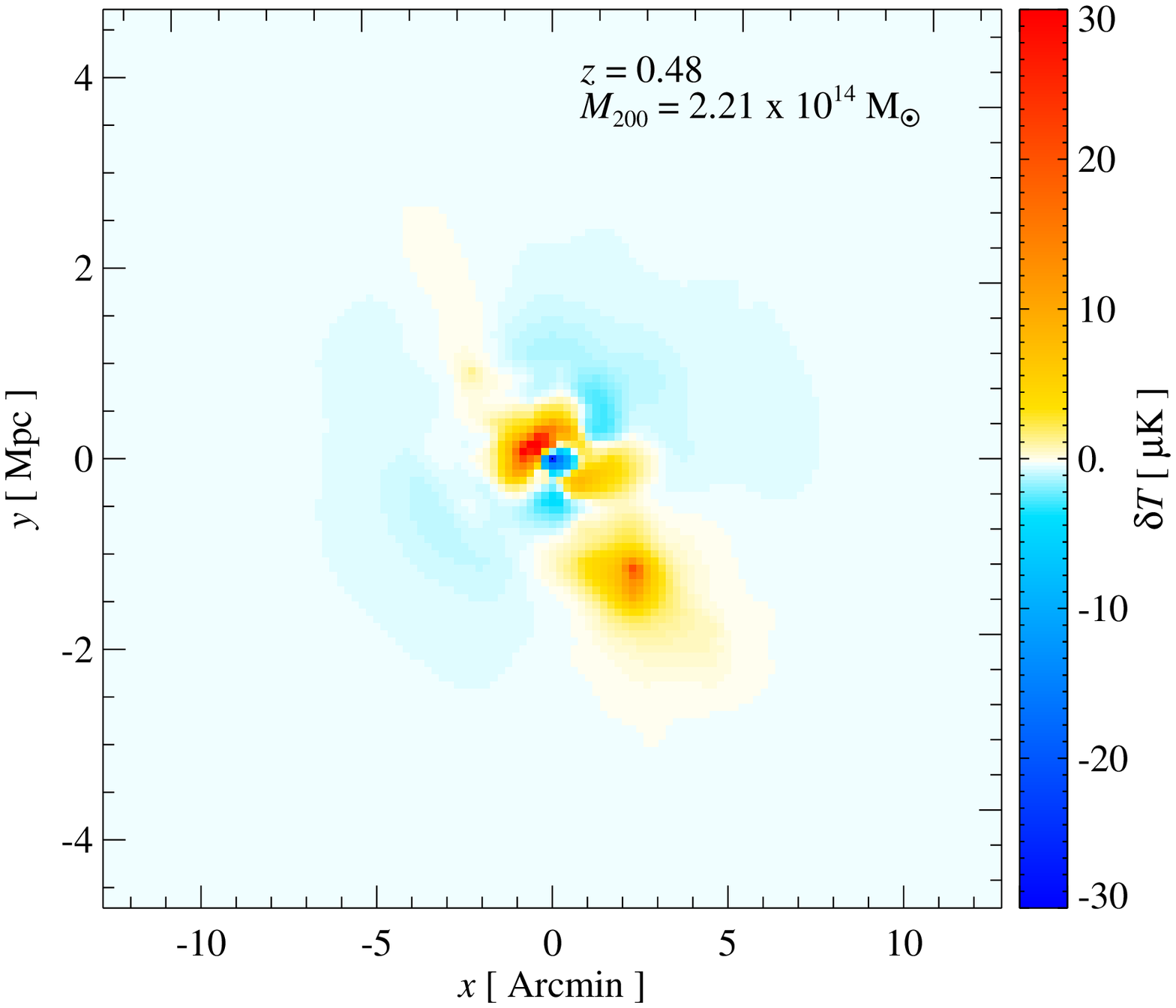}}\\
    \resizebox{0.3333\hsize}{!}{\includegraphics{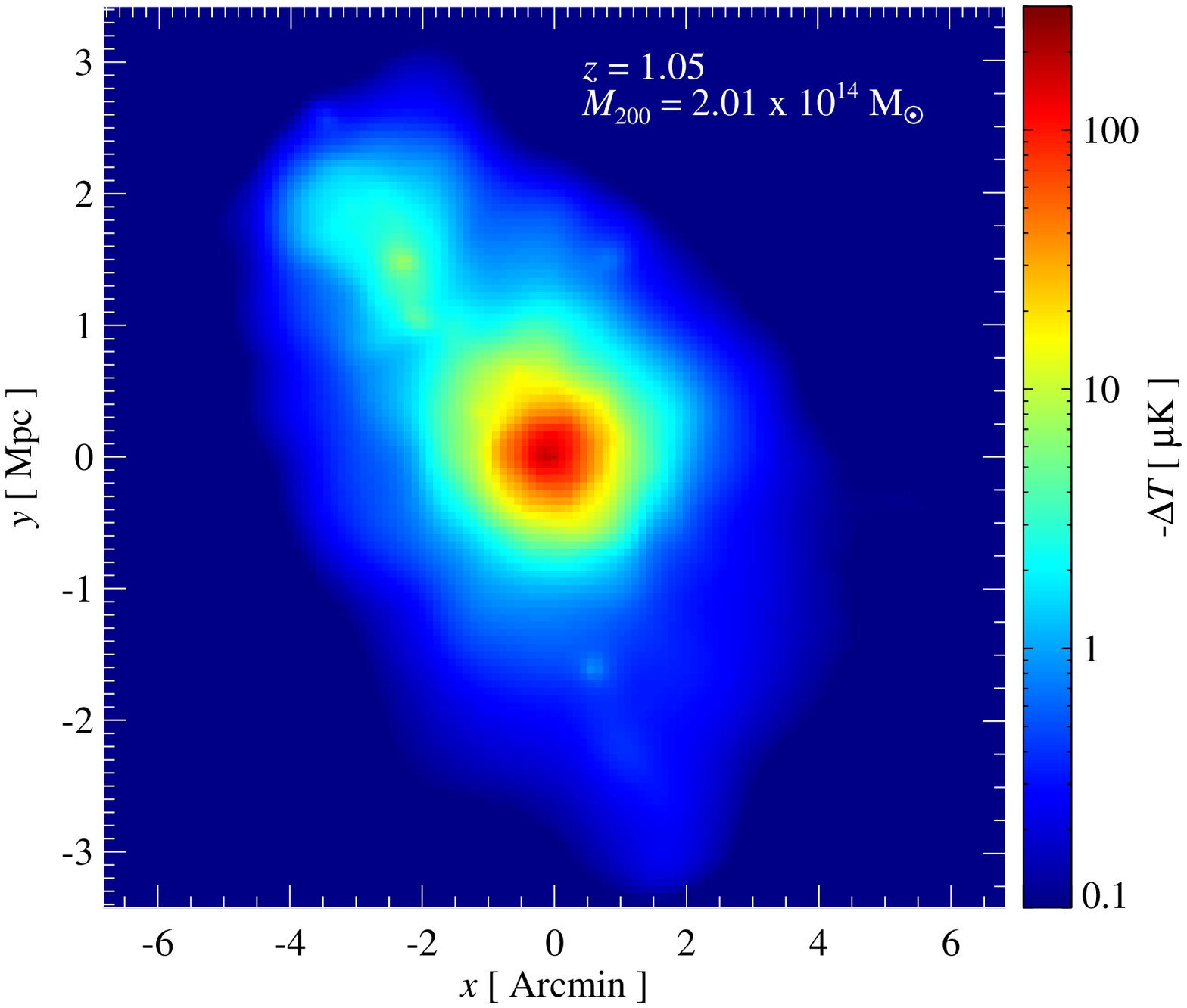}}%
    \resizebox{0.3333\hsize}{!}{\includegraphics{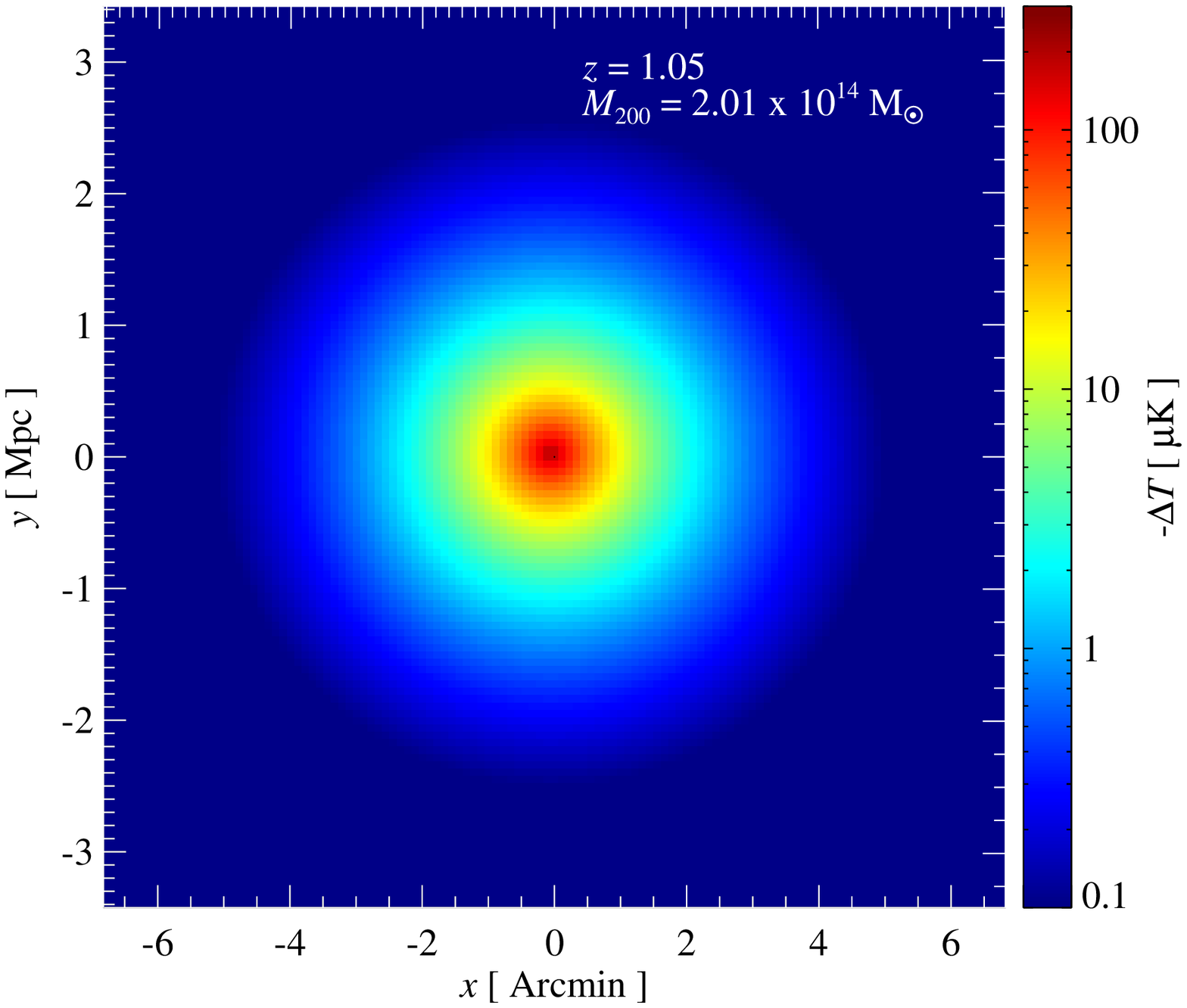}}%
    \resizebox{0.3333\hsize}{!}{\includegraphics{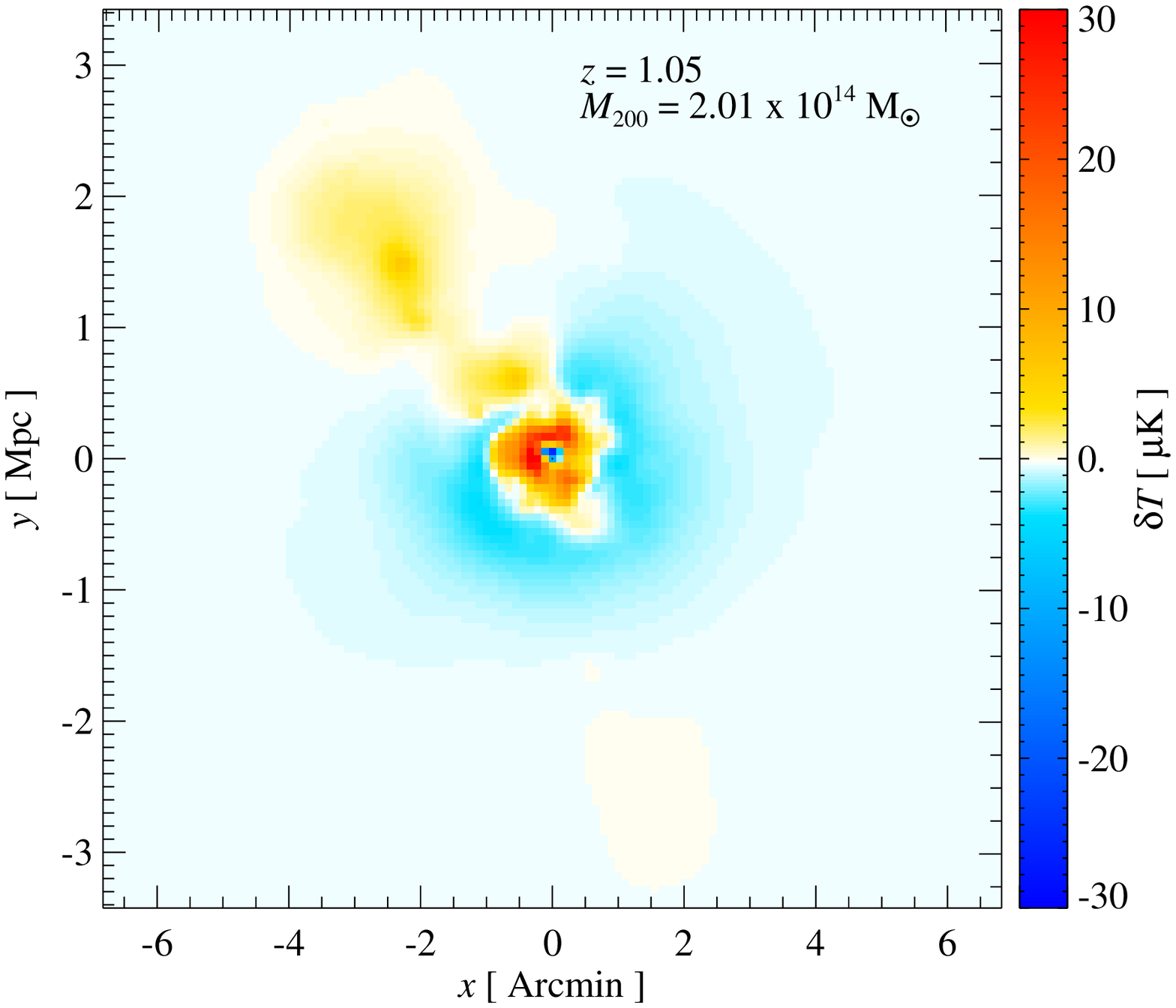}}\\
  \end{center}
  \caption[A comparison of the projected pasted profiles to four actual clusters
  from the simulations] {A comparison of four projected pressure maps of
    simulated clusters to the projected pasted-profile maps. From left to
    right, the panels show the simulated clusters (cut at a spherical radius of
    $6R_{500}$), the projected pasted profiles from the constrained fit, and the
    difference map between the two. The maps show the temperature decrement
    $-\Delta T$ in units of $\mu$K, at a frequency of 30 GHz. The difference
    maps, $\delta T$, illustrate the scales and amplitudes of the residuals
    between the simulated clusters and the projected pasted profiles. Note
    the color scale is logarithmic for the left two panels (from -0.1 $\mu$K to
    -300 $\mu$K), while it is linear for the difference map (from $-30\mu$K to
    $30\mu$K). For all panels the left and top axes are in units of Mpc and the
    bottom and right axes are in units of arc-minutes.}
\label{fig:pasteprof}
\end{figure*}

\begin{figure}[t]
\epsscale{1.2}
\plotone{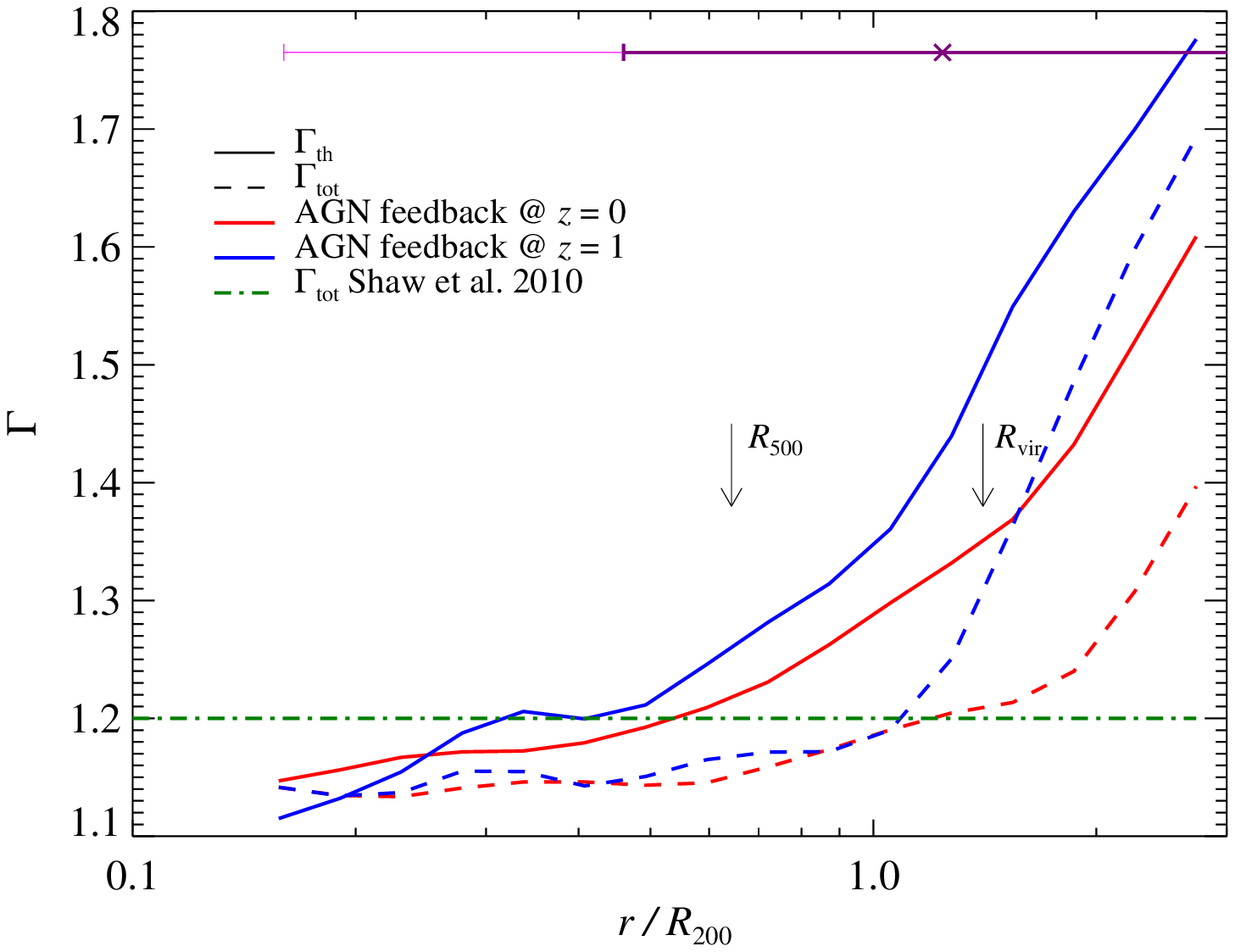}
\caption[The thermal and total adiabatic indexes as functions of
radius from simulations]{The assumption of a constant thermal or total
  logarithmic slope of the $P$--$\rho$ relation, $\Gamma_{\rmn{th}}$
  or $\Gamma_{\rmn{tot}}$, as most analytic models assume, is not
  consistent with the results from our
  simulations. $\Gamma_{\rmn{th}}$ (solid line) and
  $\Gamma_{\rmn{tot}}$ (dashed line) are shown as functions of radius
  at $z=0$ and $z=1$ from simulations with AGN feedback. For
  comparison, we show the total adiabatic index used by
  \citet{2010ApJ...725.1452S}, and we find that the differences
  increase at larger radii, especially at high redshifts.}
\label{fig:gamma}
\end{figure}

The average of this global constrained pressure profile at $z=0$
\citep[as reported in][]{2010ApJ...725...91B} compares well with the
average universal pressure profile from a representative {\em
XXM-Newton} sample of nearby systems for the region $r<R_{500}$ where
X-ray data is available \citep{2010A&A...517A..92A}. We defer the
reader to \citet{2010ApJ...725...91B} for a detailed discussion and
comparison to other numerical and observational work. However, we stress that the global constrained pressure profile of Equation
(\ref{eq:pth_mfit}) models the mass dependence and predicts a redshift
evolution that shows small but noticeable deviations from the
self-similar scaling on account of the radiative gas physics including
AGN feedback. This has clear implications when analyzing SZ
measurements of non-local clusters.

In Figure \ref{fig:pasteprof}, we present projected  30
GHz temperature maps of 4 sample clusters (cut at a spherical radius
of $6R_{500}$), their expected maps
from the global constrained fit, and the errors in the predicted temperature.  A
quantitative comparison of the tSZ power spectrum is deferred until
Section \ref{sec:clmcut}.
Hereafter, we refer to the predicted temperature maps as pasted profile maps. 
Note that this is not a representative sample of the difference
between the pasted profiles and the simulations. Instead, we attempt
to show different size clusters across different redshifts and
illustrate the scales of the deviations from the constrained fit,
primarily resulting from substructure and mis-centering, since the
cluster center of mass does not necessarily line up with the peak of
the projected pressure.  In the rightmost panel of Figure
\ref{fig:pasteprof}, we show the residuals amplitudes between the
simulated cluster projections and the pasted profile from the
constrained fits. We find that these profiles are within $\sim 10$\%
of the actual simulated cluster, which is similar to the differences
found in the bottom panels of Figure \ref{fig:pth_mass_z_fit}. These
substructures are significant on scales of tens of arc minutes for
nearby massive clusters and scales of arc minutes for higher redshift
clusters, corresponding to $\ell \sim 1000 - 10000$.

\subsection{Analytic Assumptions in the Thermal Pressure Profile}

Analytic and semi-analytic models typically rely on assuming an
pressure-density ($P$--$\rho$) relation and some form of hydrostatic
equilibrium (HSE), possibly including non-thermal support terms.
Fully analytic models,
\citep[e.g.,][]{2002MNRAS.336.1256K,2010ApJ...725.1452S}, apply HSE to
theoretical, spherically symmetric dark matter potentials.
Semi-analytic models,
\citep[e.g.,][]{2010ApJ...709..920S,2011ApJ...727...94T}, take dark
matter simulations, and paste baryons on top of the dark matter
potential wells, again using (possibly corrected) HSE and an
$P$--$\rho$ relation.  The results from both classes of models, then,
rely critically on the input $P$--$\rho$ relation and are sensitive to
departures from HSE.  In contrast, empirical fits to the average
cluster pressure profile derived from simulations have a key advantage
over analytical models because the simulations naturally account for
kinetic pressure support from non-thermalized bulk flows which provide
substantial support in the outer parts of clusters, but do not
contribute to the tSZ.  They also make no assumptions about HSE (which
is grossly violated during, for instance, mergers), and rather than
forcing an $P$--$\rho$ relation, they track the flow of energy into
and out of the ICM.

The (semi-)analytic calculations cast the $P$--$\rho$ relation in
terms of a pressure law $P\propto \rho^{\Gamma}$, and usually assume a
constant $\Gamma$, where $P$ can be either the thermal pressure
$P_{\rmn{th}}$ which is the source for the tSZ effect, or the total
pressure, $P_{\rmn{tot}} \equiv P_{\rmn{th}} + P_{\rmn{nt}}$, where
$P_{\rmn{nt}}$ is any non-thermal support, principally kinetic motion
of the ICM\footnote{Some older models have ignored kinetic support
  entirely, in which case $P_{\rmn{tot}} = P_{\rmn{th}}$.}.  The total
pressure is the input to the equation of hydrostatic equilibrium that
reads for spherical symmetry

\begin{equation} \label{eq:HSE}
\dd P_{\rmn{tot}}\,/\,\dd r = -GM(<r)\rho\,/\,r^2.
\end{equation}

We present the effective adiabatic index, $\Gamma=\dd\log P/\dd\log\rho$, as a
function of cluster radius in Figure \ref{fig:gamma}.  We find that the
assumption of constant $\Gamma$ is grossly violated, particularly in the outer
parts of clusters, and for $P_{\rmn{th}}$.  These results stress the importance
of deriving pressure profiles from observations and hydrodynamical simulations,
particularly as good-quality observational data from cluster outskirts is in
short supply.

\section{The tSZ power spectrum in detail}
\label{sec:PS}

\begin{figure*}
  \resizebox{0.5\hsize}{!}{\includegraphics{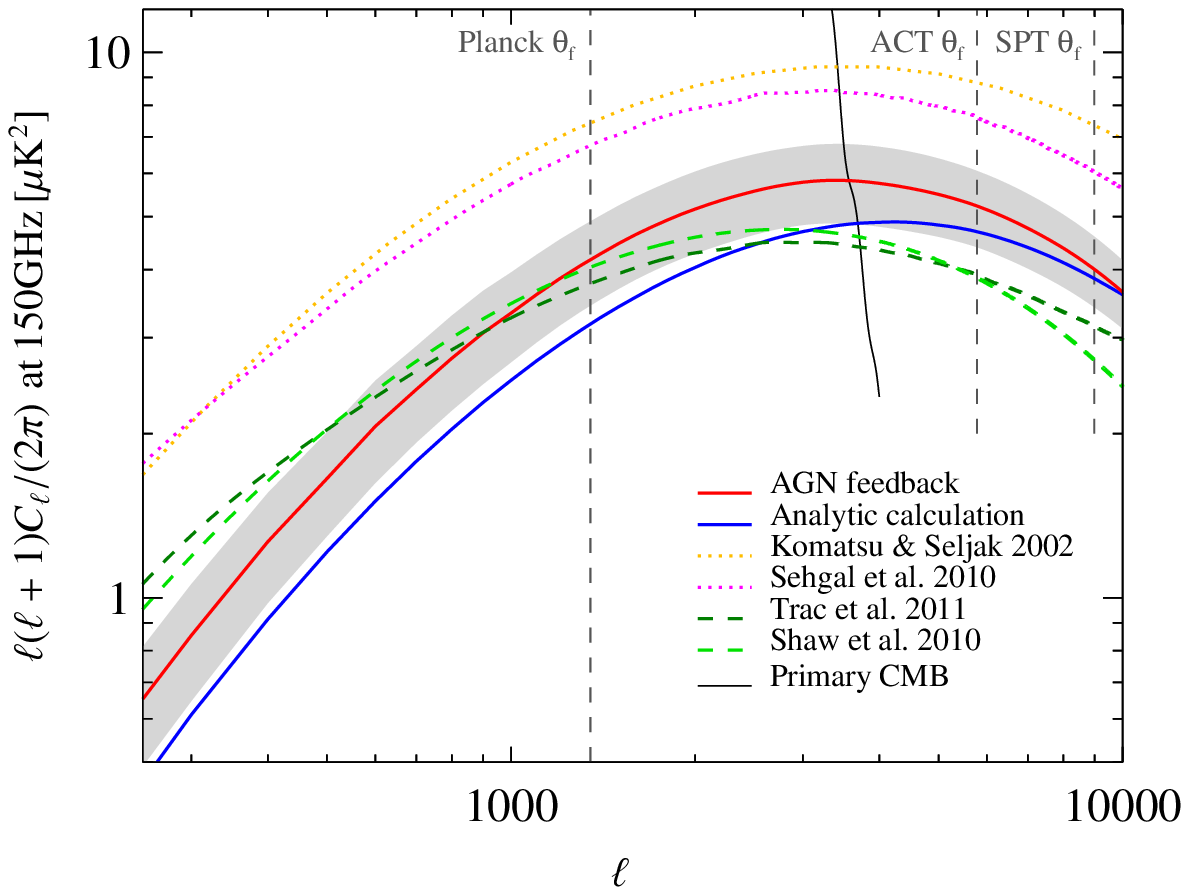}}%
  \resizebox{0.5\hsize}{!}{\includegraphics{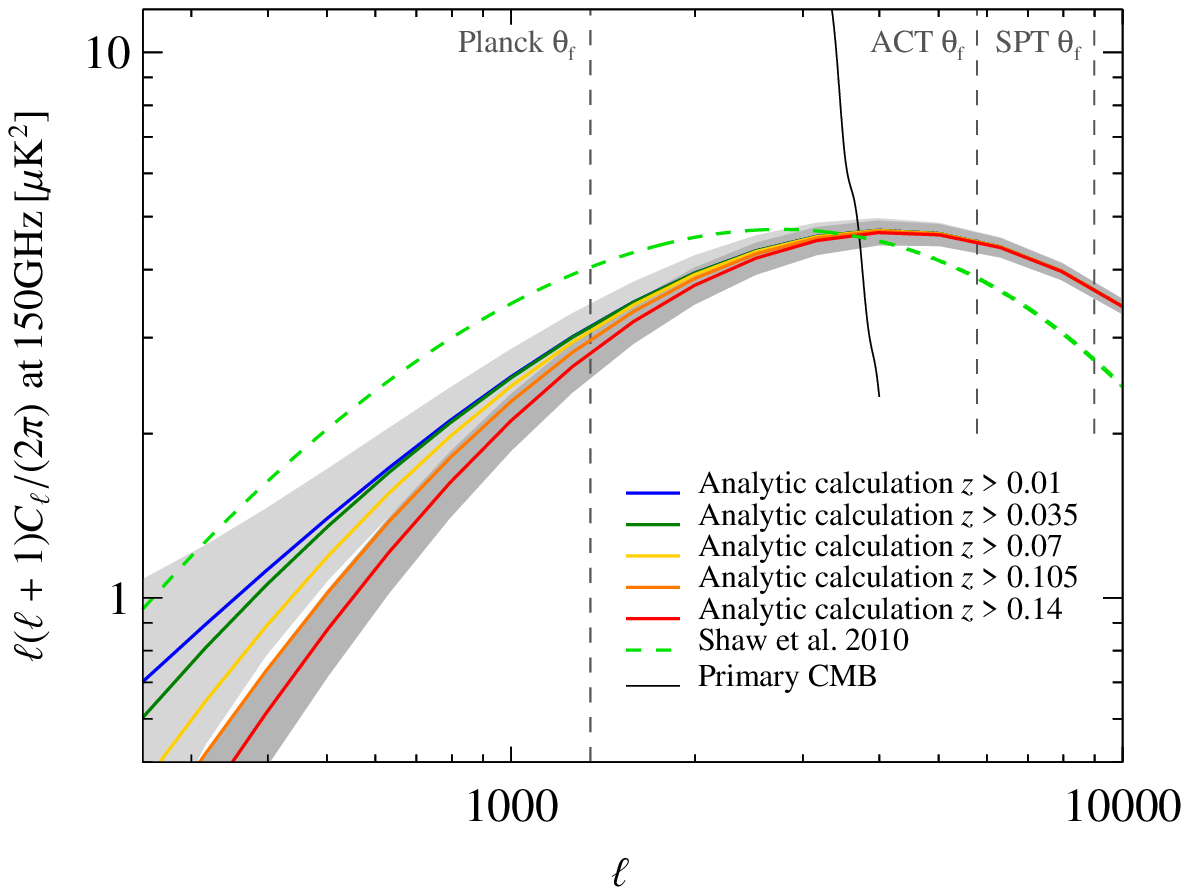}}\\
  \caption[Comparison between the current predictions for the thermal Sunyaev-Zel'dovich power spectra at 150 GHz]{In the left panel, we show a comparison
    of the current predictions for the tSZ power spectra at 150 GHz from our
    simulations with AGN feedback (red line) and the analytical calculations
    using the constrained pressure profiles in this work (blue line). The
    standard deviation among our 10 simulations is shown with a light grey band.
    We also include the semi-analytical simulations by
    \citet{2010ApJ...709..920S} (pink dotted line) and
    \citet{2011ApJ...727...94T} which includes enhanced non-thermal pressure
    support (dark green dashed line) and the fully analytical calculations by
    \citet{2002MNRAS.336.1256K} (orange dotted line) and
    \citet{2010ApJ...725.1452S} (light green dashed line). The full-width
    half-max values appropriate for the Planck, ACT and SPT beams are also
    plotted.  At low-$\ell$, our two methods of calculating the tSZ diverge
    because our simulations happen to contain a large number of high mass
    objects driving the power up, though the excess is consistent with expected
    Poisson fluctuations.  At high-$\ell$ the discrepancy is the result of
    substructure and asphericity, as demonstrated in Sections \ref{sec:clmcut}
    and \ref{sec:clzcut}. The right panel shows a comparison between the current
    analytic calculations for the tSZ power spectra and how the power spectrum
    changes with the variation of the lower redshift limit of integration. The
    variance of 1\% of the full-sky power spectrum (cf. Equation~(\ref{eq:clerr})) is
    illustrated by the grey bands for the highest and the lowest redshift limits
    of integration.}
 \label{fig:clcomp}
\end{figure*}

In this section we compare three different ways of calculating the tSZ power
spectrum: directly projecting the electron pressure in the simulations, taking
the simulation cluster catalogs and projecting our constrained global pressure
profile onto the cluster locations (the ``pasted-profile'' maps), and using a
completely analytical halo calculation.  For the analytic calculation, we use
the formalism described in Section \ref{sec:tSZthry} and the constrained
pressure profile from Section \ref{sec:prof}.  For the simulation and
pasted-profile approaches, the thermal Compton-$y$ maps are obtained by
performing a line-of-sight integration of the electron pressure through the
entire simulation box at each redshift output, covering $z=0.07$ to $z=5$. For
each redshift-output map we compute the average power spectrum for our ten
simulations and add these differential power spectra up\footnote{We have
  selected the redshifts at which we write out the simulation snapshots to be
  the light crossing time of the simulation; hence, the total power spectrum is
  the sum of the differential power spectra.}.  This procedure uses all the
information within the simulation volume and decreases the variance of the power
spectrum, especially at low redshifts.  One benefit of this technique is that by
summing over redshift slices {\it after} taking the power spectra, we ignore any
correlations between different redshift slices, as effectively happens in
nature.  With more traditional methods that stack redshift slices (such as were
used in \citealt{2010ApJ...725...91B}), care must be taken that different
redshift slices do not project the same objects to the same locations, as that
induces artificial correlations, potentially altering the tSZ power spectrum.

In the left panel of Figure \ref{fig:clcomp}, we plot the tSZ power
from our analytical halo calculation and that from the AGN
simulations.  For reference, we include other tSZ power spectrum
templates
\citep{2002MNRAS.336.1256K,2010ApJ...709..920S,2010ApJ...725.1452S,2011ApJ...727...94T}. We
choose the cosmological parameters for the halo calculation to match
the simulations and integrate from $z=0.07$ to $z=5$, so that the only
possible sources of differences are the mass function and the pressure
profiles. There are clear differences between the analytical halo
calculation and the complete simulation maps. The main difference at
low $\ell$s results from shot noise within the sample of simulated
boxes, where we had more (though consistent within the expected error)
high-mass clusters than expected, but this is only a 6\% effect in the
total power spectrum (cf. Appendix).  The differences at higher
$\ell$s arise from deviations about the average pressure profile,
including effects of cluster substructure and asphericity. We see these
variations in the residual maps of individual simulated cluster
projections and pasted profile projections
(cf. Fig. \ref{fig:pasteprof}). We further explore these differences
in the power spectrum between the analytic calculation and the
simulations in the following Sections \ref{sec:clmcut} and
\ref{sec:clzcut}. It is challenging to determine the causes for all
the differences between our calculations and other calculations for
the tSZ power spectrum
\citep{2002MNRAS.336.1256K,2010ApJ...709..920S,2010ApJ...725.1452S,2011ApJ...727...94T},
since the thermal pressure profile we use is different from the ones used
by the other calculations. However, the reasons for the differences we
find between our three methods, will be generally applicable to the
other methods of calculating the tSZ power spectra.

The right panel of Figure \ref{fig:clcomp} shows a direct comparison between our
analytical model and the \citet{2010ApJ...725.1452S} model. In both
calculations, the same cluster mass function was used and the power spectra are
scaled to the same cosmological parameters, so the differences are related to
the model for the thermal pressure profile.  We investigate the redshift
integration limits\footnote{For the remainder of this paper, we use a low
  redshift cutoff of $z = 0.07$, so that we can directly compare our analytic
  calculation to the simulations.}, but find they do not significantly affect
the differences at $\ell \gtrsim 1000$.  We present the expected mean and
standard deviation of 1\% of a full-sky tSZ measurement as a function of lower redshift
cutoff, and find that the low-$\ell$ variance is substantially suppressed by
raising the low-$z$ cutoff.  On the scales where the tSZ peaks, we find both the
mean spectrum and the variance are only weakly affected by varying the redshift
limit from $z$=0.01 to $z$=0.14. Similar results have been found when making
intensity cuts on sky maps \citep{2009ApJ...702..368S}.

We now present power spectra calculated directly from the simulations.  In
addition to projecting the full electron pressure from all
particles, we also take advantage of the information from the
simulation cluster catalogs.  By doing this, we can employ mass,
redshift, and radius cuts to explore the dependence of the full tSZ
power spectrum.  By pasting our global pressure profile to locations
and redshifts of simulated clusters, we can also explore, without having
to worry about sample variance, the effects of using our profile
instead of the full simulation results.

We use the cluster catalogs described in Section \ref{sec:pfit}, and
remind the reader that $M_{FOF}$ is roughly equal to $M_{200}$, though
with large scatter.\footnote{For detailed work on comparing the mass
  definitions of $M_{\rmn{FOF}}$ to $M_{\Delta}$ and the resulting halo
  mass catalogs from these definitions see \citet{2011ApJS..195....4M}
  and the references therein.}
Our cluster mass function becomes incomplete below $M_{200} \sim 4
\times 10^{13} M_{\sun}$ (cf. Appendix) primarily due to our $M_{FOF}$ cutoff in the
original cluster finding of $1.4 \times 10^{13} M_{\sun}$, but
partially due to the linking length merging some clusters/groups into
nearby larger clusters at the $10-15$\% level
\citep[e.g.,][]{1985ApJ...292..371D,1991ComPh...5..164B,1996MNRAS.281..716C,2008MNRAS.385.2025C}.
For these reasons we examine  only cluster with
$M_{500} > 4.2\times10^{13}\,M_{\sun}$ when we bin clusters in mass.   

\begin{figure}
  \epsscale{1.2}
  \plotone{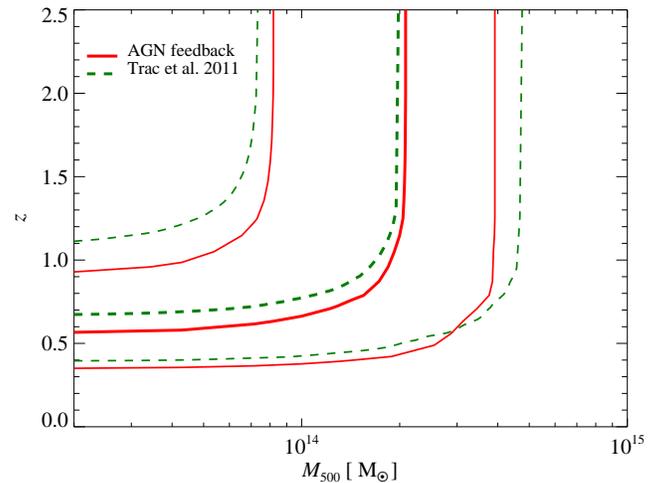}
  \caption[The cumulative distribution function the thermal Sunyaev-Zel'dovich power spectrum as a function of mass and redshift at $\ell
    = 3000$]{Shown is the cumulative distribution function for the
    thermal SZ power spectrum as a function of mass and redshift at $\ell
    = 3000$.  The curves show the lower mass and upper redshift
    cutoffs that contribute [25,~50,~75]\% to the tSZ power spectrum.
    At $\ell = 3000$, half the power of tSZ power spectrum comes
    from clusters with $z > 0.6$, and half comes from clusters
    with $M_{500} < 2\times10^{14} M_{\sun}$.  For comparison, the
    dashed green lines show the semi-analytical results of
    \citet{2011ApJ...727...94T}, which include enhanced non-thermal
    pressure support.}
  \label{fig:clcdf}
\end{figure}

In Figure \ref{fig:clcdf} we show the cumulative distribution function (CDF) for
the tSZ power for a CDF$(M> , z<)$ at $\ell = 3000$. The CDF illustrates where
the relative amount of power originates at the 25\%, 50\% and 75\% percentile
levels. Half the power at $\ell = 3000$ comes from clusters with $z > 0.6$ and
half originates from clusters with mass $M_{500} < 2\times10^{14}
M_{\sun}$. This result is in general agreement with other work
\citep{2002MNRAS.336.1256K,2011ApJ...727...94T}. We note that the particulars of
these mass and redshift ranges are sensitive to the input modeling of the
ICM. The comparatively low mass and high redshift of the clusters and groups
that make up the bulk of the tSZ signal mean that they have not been as well
studied as more massive and nearby objects.  Thus, the tSZ angular power
spectrum can provide a statistical constraint on the astrophysical processes of
importance at high redshift and in low-mass clusters.

\subsection{Contribution to the tSZ Power Spectrum in Cluster Mass Bins}
\label{sec:clmcut}

\begin{figure*}
  \resizebox{0.5\hsize}{!}{\includegraphics{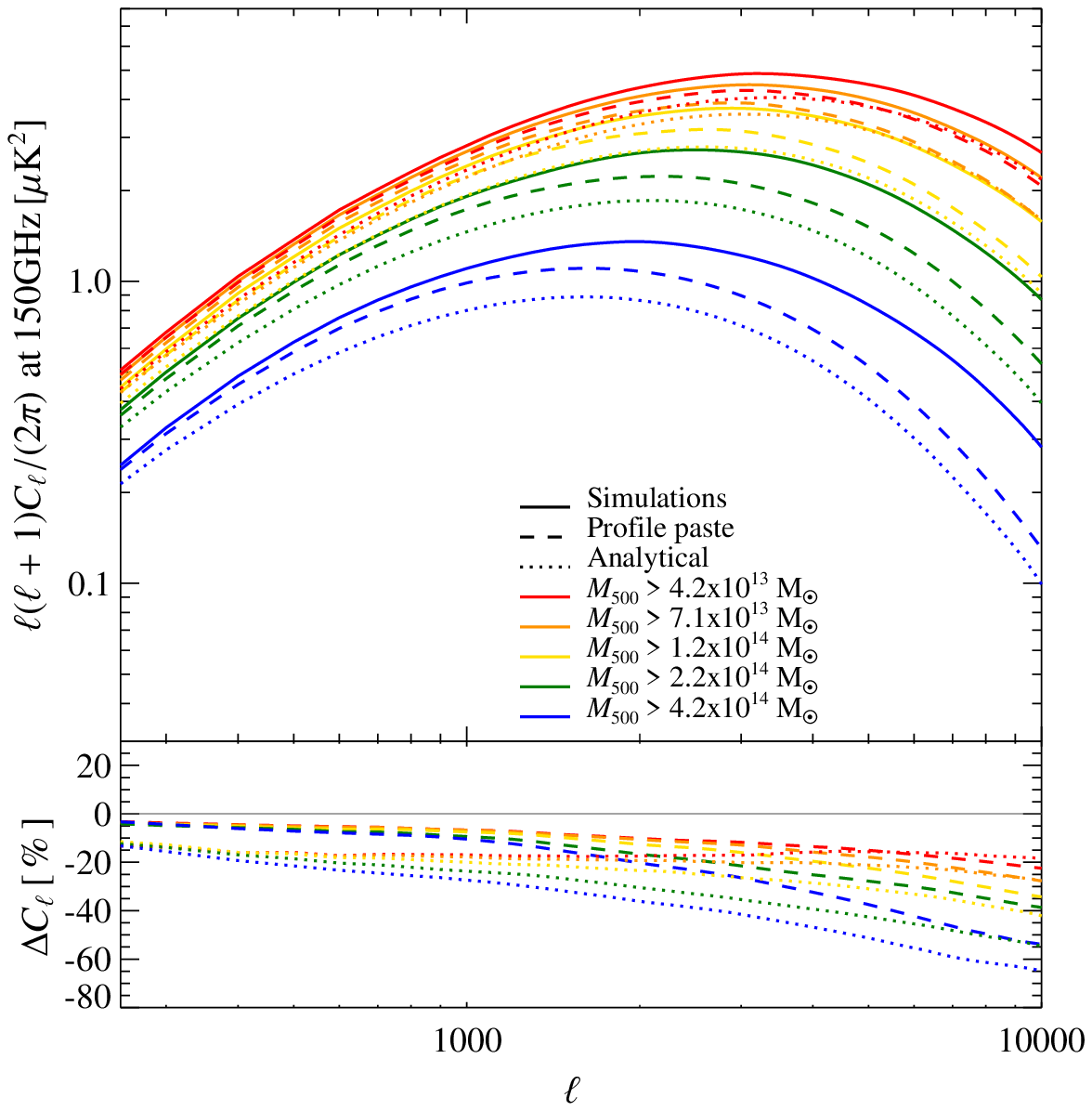}}%
  \resizebox{0.5\hsize}{!}{\includegraphics{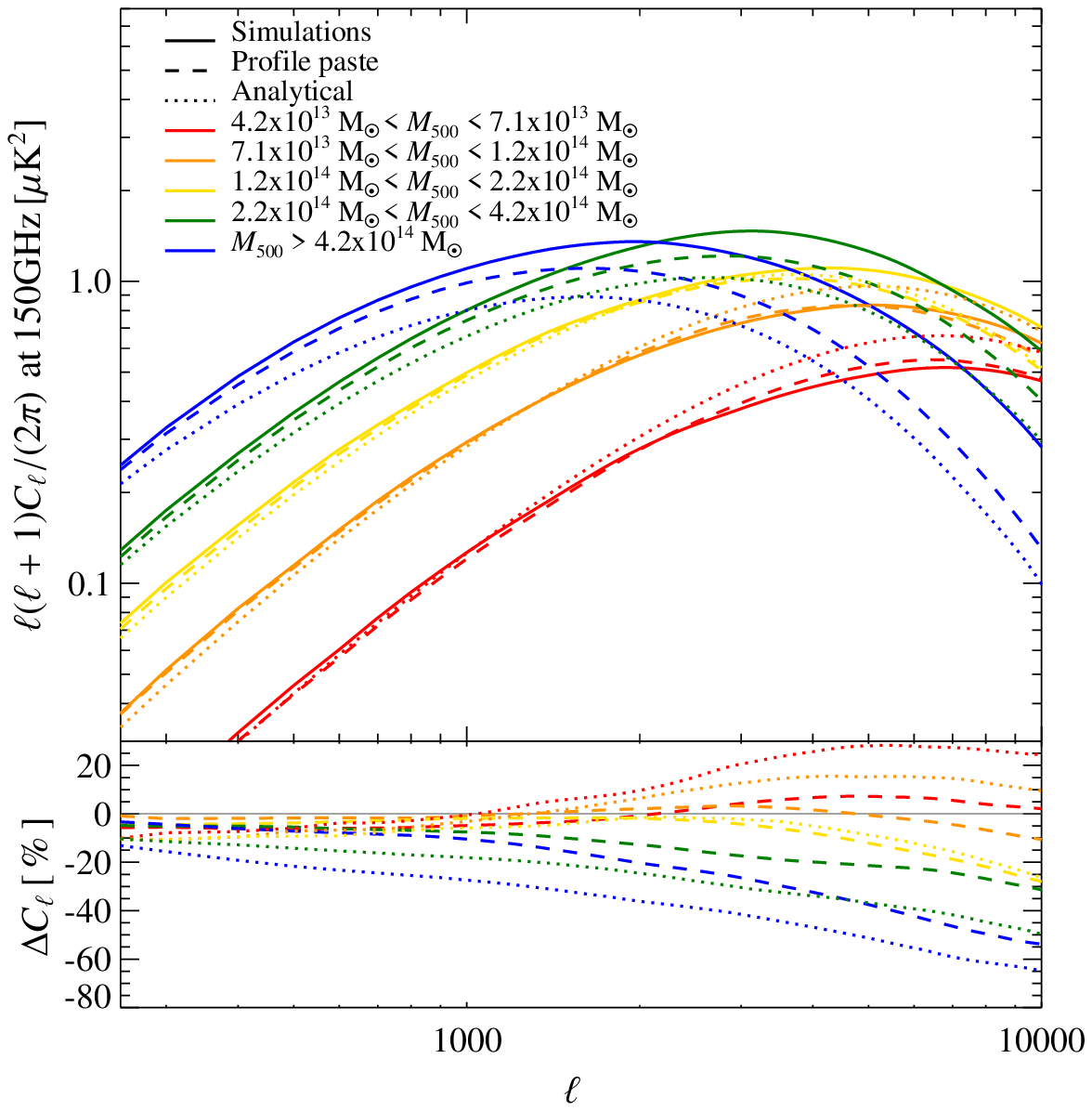}}\\
  \caption[The thermal Sunyaev-Zel'dovich power spectrum sorted into bins of
  galaxy cluster mass]{The tSZ power spectrum sorted into bins of cluster
    mass. Left: we show the cumulative tSZ power spectrum in mass bins
    ($\cltsz\,(M_{500}\,>\,M_{\mathrm{cut}}$) from the AGN feedback simulations,
    the pasted profile maps and the analytical calculation. Right: we show the
    differential tSZ power spectrum
    $\cltsz\,(M_{\mathrm{cut,low}}\,<\,M_{500}\,<\,M_{\mathrm{cut,high}}$) for
    the same power spectrum calculations. In the bottom of both panels we show
    the relative difference, $\Delta C_{\ell} = 100\left(C_{\ell,sim} -
      C_{\ell,i}\right)/ C_{\ell,sim}$, where $C_{\ell,sim}$ is the power
    spectrum of the simulated maps and $C_{\ell,i}$ is that from the pasted
    profile maps and the analytical calculation. The differences between the
    simulations and the pasted profile maps result from the absence of
    substructure and asphericity in the pasted profile maps, which is larger for more massive clusters. The larger
    differences found between the analytical calculation and the simulations are
    the result of the mass catalog of the simulations having an excess of high
    mass clusters and deficit of lower mass cluster compared to the analytic
    mass function (cf. Fig. \ref{fig:mf}).}
\label{fig:clmcut}
\end{figure*}

In this subsection, we calculate the power spectrum in mass bins.  This allows
us to isolate the differences between the simulations, the pasted profile maps,
and the analytic calculation, as functions of cluster mass, integrating in
redshift between $z =0.07$ and $z = 5$.  We explore both, cumulative and
differential mass bins.  We consider all gas particles (or radii) within $6
R_{500}$ when projecting the thermal pressure of the simulations. Our method
takes care of not double counting the cluster mass in overlapping volumes of
close-by clusters.  In Figure \ref{fig:clmcut}, we show the power spectrum
broken down into cumulative (left panel) and differential (right panel) mass
bins.  The bottom panels show the relative differences, where $\Delta C_{\ell} =
100 \left(C_{\ell,sim} - C_{\ell,i}\right)/ C_{\ell,sim}$, with $C_{\ell,sim}$
denoting the power spectra from the simulations and the $C_{\ell,i}$ are the
power spectra from either the projected pasted profile maps or the analytic
calculation.

The largest deviations between our analytic/pasted profile spectra and the full
simulations are for the highest mass ($M_{500} \gtrsim 4.2 \times 10^{14}
M_{\sun}$) clusters, particularly on small angular scales.  The deviations
between the pasted profiles and the simulations in this mass range arise from
the increased level of substructure and asphericity in massive clusters in
comparison to smaller objects due to the more recent formation epoch of large
systems in a hierarchical structure formation \citep{Wechsler+2002,Zhao+2009,
  PCB,battinprep}.  The high-mass difference between the fully analytic tSZ
spectrum and the simulation results reflects our overabundance of high-mass
clusters due to shot noise relative to the mass function used in the analytic
calculation. The agreement between all three methods is excellent for masses
below $ 4.2\times 10^{14} M_{\sun}$ until our cluster catalog becomes incomplete at
low masses.  In the most massive cluster bin, the relative differences between the power
spectra are $\sim40 - 60$\% for $\ell\sim 2000 - 9000$ (cf. Fig
\ref{fig:clmcut}). For the lower mass bins the differences fluctuate between
$\pm 30$\%, with the pasted profiles generally agreeing better with the full
simulation results.

\subsection{Contribution to the tSZ Power Spectrum in Redshift Bins}
\label{sec:clzcut}

\begin{figure*}
  \resizebox{0.5\hsize}{!}{\includegraphics{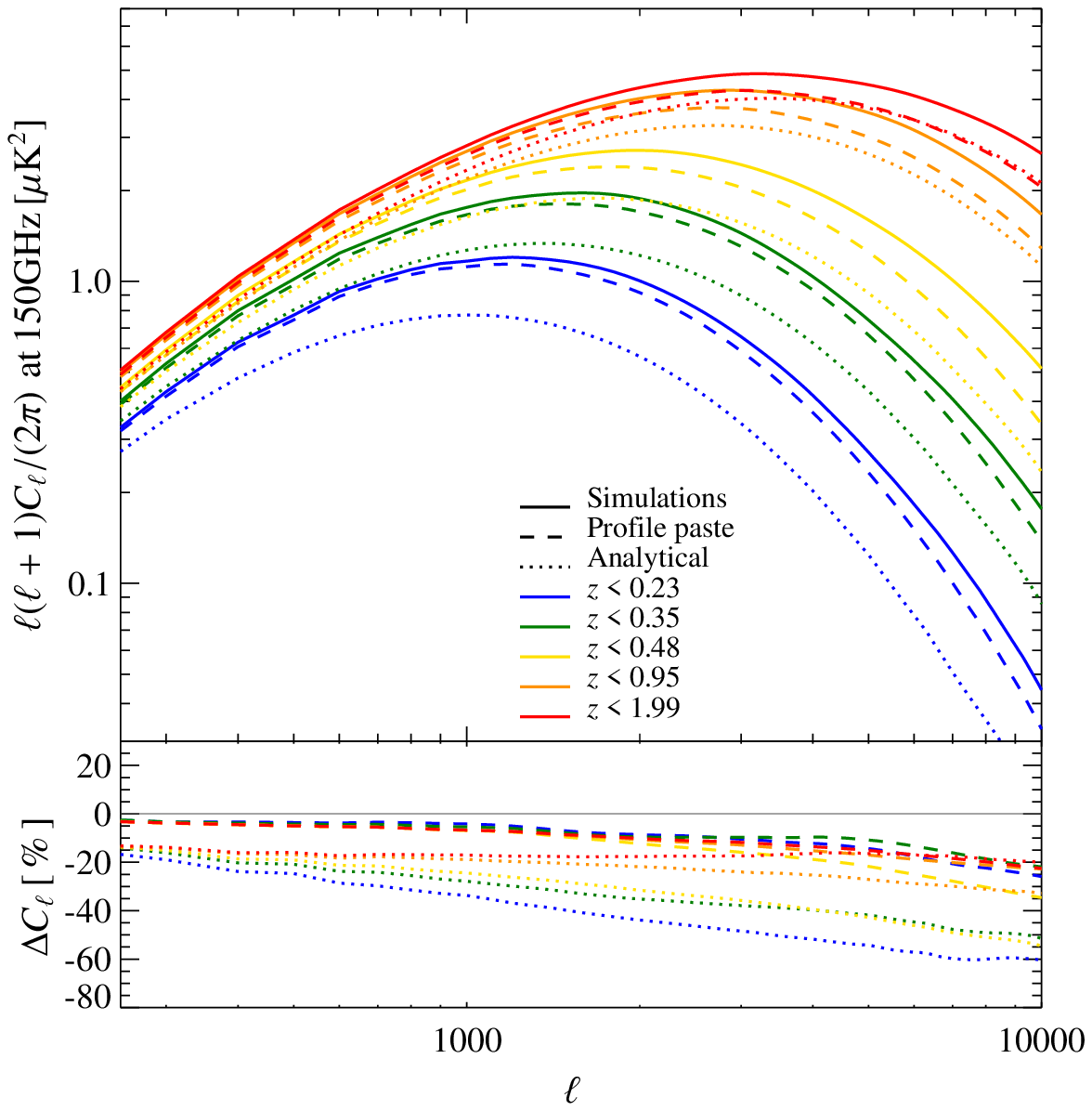}}%
  \resizebox{0.5\hsize}{!}{\includegraphics{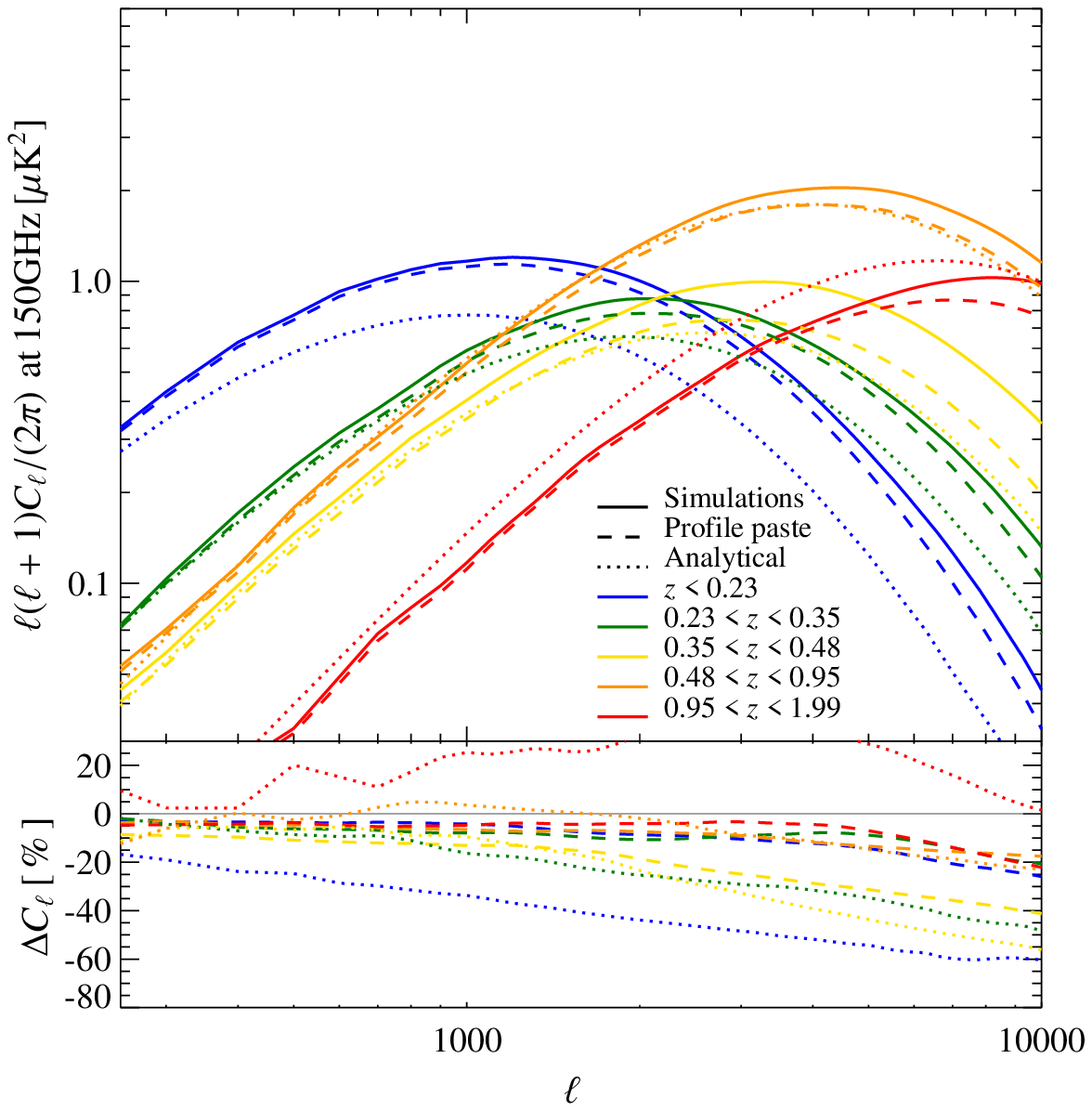}}\\
\caption[The thermal Sunyaev-Zel'dovich power spectrum sorted into
redshift bins]{The same as Figure \ref{fig:clmcut}, however for
redshift slices.  Left: we show the cumulative tSZ power spectrum in
redshift bins $\cltsz\,(z\,<\,z_{\mathrm{cut}})$ from the AGN feedback
simulations, the pasted profile maps and the analytical
calculation. Right: we show the differential tSZ power spectrum
$\cltsz\,(z_{\mathrm{cut,low}}\,<\,z\,<\,z_{\mathrm{cut,high}})$ for
the same power spectrum calculations. In the bottom of both panels we
show the relative difference, $\Delta C_{\ell} = 100\left(C_{\ell,sim}
- C_{\ell,i}\right)/ C_{\ell,sim}$, where $C_{\ell,sim}$ is the power
spectrum of the simulated maps and $C_{\ell,i}$ is that from the
pasted-profile maps and the analytical calculation.  The agreement
between the pasted profile and simulation spectra is excellent below
$\ell \sim 5000$ for all redshifts.  On smaller scales, cluster
substructure contributes similarly across all redshift bins examined.
}
\label{fig:clzcut}
\end{figure*}

In this subsection we calculate the power spectrum in redshift bins and compare
the results from the simulation, the pasted profile maps, and the analytical
calculation to aid in understanding the differences between these approaches. In
Figure \ref{fig:clzcut}, we show the power spectrum broken down into cumulative
(left panel) and differential (right panel) redshift bins.  Here we fix the mass
range to $M_{500} > 4.2 \times 10^{13} M_{\sun}$ and set the lower redshift
integration bound for the cumulative spectra to $z = 0.07$.  We use the same
definition for $\Delta C_{\ell}$ to show the differences between power spectrum
calculations. In contrast to the mass cuts, the differences between the
projected simulated maps and the pasted-profile maps are similar across all the
redshift slices (cf. Fig. \ref{fig:clzcut}). For $\ell < 5000$, there is a $\sim
5-10$\% difference between the pasted profiles and the simulations, rising to
$\sim 20$\% at $\ell=10,000$.  These results suggests that the contributions from
substructure and asphericity to the power spectrum are similar across the
redshift range explored, with the exception of one redshift bin $z\sim 0.4$
which contains a rare merger event.  The large deviations between the analytic
and simulation/profile-paste spectra in the highest redshift bin are likely due
to the incompleteness of the cluster catalogs at the lowest masses, which are
preferentially more important at high redshift.  At low redshift, we attribute
the difference between the analytic and the profile-paste power spectra to the
shot noise in the most massive clusters.

\subsection{Contribution to the tSZ Power Spectrum within given Cluster Radii}
\label{sec:clrcut}

\begin{figure}
\epsscale{1.2}
\plotone{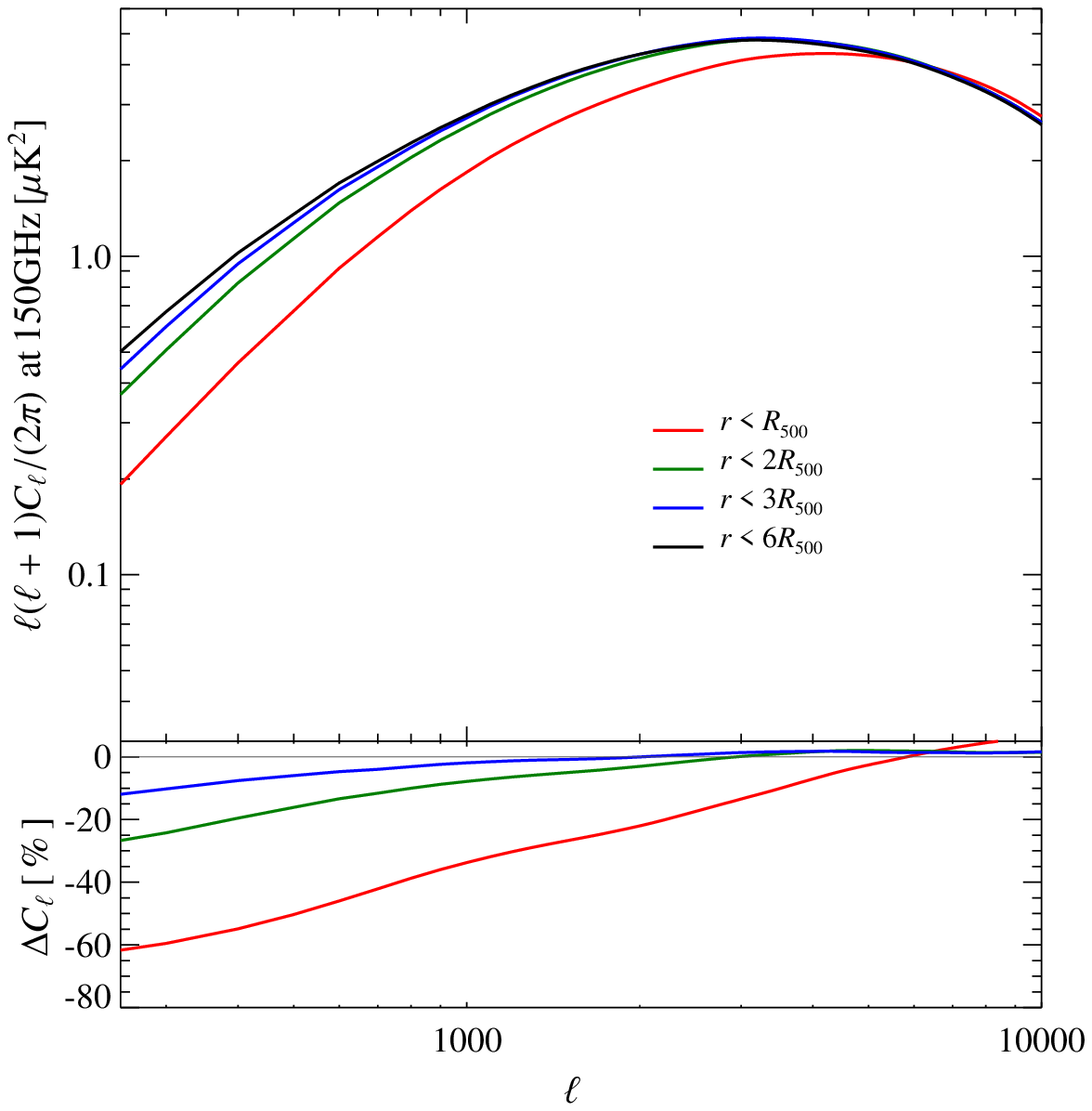}
\caption[The thermal Sunyaev-Zel'dovich power spectrum for a given radial
truncation] {$\cltsz\,(r\,<\,R_{\mathrm{cut}}$) for the AGN feedback
simulations.  The thermal pressure distribution
has been tapered as in Equation~(\ref{eq:taper}) at varying cluster-centric radii before projection.
On small scales, virtually all of the power at $\ell > 2000$ comes
from $r<2R_{500}$.  About 80\% of the tSZ power is recovered at
$\ell=3000$ when tapering at $R_{500}$, though the deviations become
substantially larger at smaller $\ell$.  These results emphasize the
importance of understanding cluster pressure profiles well past
$R_{500}$ in order to do high-precision work with the tSZ power spectrum.}
\label{fig:clradcut}
\end{figure}

In this subsection we apply radial truncations to the full simulated
pressure maps, using clusters with $M_{500} > 4.2 \times 10^{13} M_{\sun}$
at $0.07 < z <5$. The procedures for making real space radius cuts in
maps or analytical calculations are not trivial, since any sharp cut
in real space produces ringing in Fourier space, potentially transferring
power from large to small angular scales. 
To reduce ringing and the potential to artificially increase the high-$\ell$
power spectrum, we use a Gaussian taper when truncating the pressure profile.
We place radial tapers at $r = R_{500}$, $2R_{500}$, $3R_{500}$, and $6R_{500}$
in the maps, adopting $6R_{500}$ as the reference radial taper\footnote{We avoid
  double counting gas particles when we project them into maps. If a particle
  lies in the overlap region between two clusters, we taper the particle with
  the larger of the two possible taper values, i.e. those particles with a
  smaller radius $R/R_{500}$, to avoid artificially suppressing power in the overlap
  region.}.  The form of the taper is
\begin{equation}
\mathcal{T}(r)=\exp\left [-\left ( \frac{r-r_t}{80 \times R_{500}}\right  )^2 \right
]
\label{eq:taper}
\end{equation}
for $r$ greater than the taper radius $r_t$, and unity otherwise.  In
the bottom panel of Figure \ref{fig:clradcut} we show the relative
difference, $\Delta C_{\ell} = 100\left(C_{\ell,6R_{500}} -
C_{\ell,i}\right)/ C_{\ell,6R_{500}}$, where $C_{\ell,6R_{500}}$ is
the power spectrum from the $6R_{500}$ radial cut and $C_{\ell,i}$ are
power spectra from the other radial cuts.  The trend we find is that
the large radii of clusters are only important for the low $\ell$s,
for example the contributions to the tSZ power spectrum when only
integrating out to $R_{500}$ yields $\sim30 - 65$\% of the total power
from $\ell = 100 - 1000$, respectively. At $\ell = 3000$ only about
10\% of the total tSZ power comes from beyond $R_{500}$. This number
is consistent with previously quoted values
\citep{2011ApJ...727L..49S}. We note that there is some small residual
Fourier ringing, as the tapered spectra rise above the fiducial at
$\ell$s of many thousand.  Nevertheless, at higher $\ell$, the cluster
centers begin to be resolved and become the dominant contributors to
the tSZ spectrum since their surface brightnesses are so much larger
than any emission in the cluster outskirts.

\section{Constraints of $\sigma_8$ from Current ACT and SPT Data}
\label{sec:sig8}

Using the tSZ power spectrum and ignoring any template uncertainty, the
constraints on $\sigma_8$ are competitive with other cosmological measurements.
After accounting for template uncertainty, there is no statistically significant
discrepancy between $\sigma_8$ determined from the tSZ power and that derived
from primary CMB anisotropies, or other the measurements
\citep{2010arXiv1009.0866D,2010arXiv1012.4788S}. Here we use our
$C_{\ell,\mathrm{tSZ}}$ templates at the fiducial parameters $\sigma_8=0.8$ (and
$\Omega_{\rmn{b}}h=0.03096$) to define the shape of the tSZ power spectrum, and
content ourselves with determining only the template amplitude, $\atsz$,
relative to that expected from the background cosmology
\citep[e.g.,][]{2010ApJ...725...91B,2010arXiv1009.0866D}. The amplitude of
$\atsz$ is proportional to a large power of $\sigma_8$ \citep[$\atsz \propto
\sigma_8^{7\ldots
  9}$][]{2002ASPC..257...15B,2002MNRAS.336.1256K,2005ApJ...626...12B,2011ApJ...727...94T}. It
follows that values of $\atsz$ below unity imply that theoretical templates
overestimate the SZ signal, or else points to a smaller value of $\sigma_{8}$
than the value derived from primary CMB anisotropies.

The probability distributions of the amplitude, $\atsz$, and other cosmological
parameters are determined from current CMB data using a modified version of
CosmoMC \citep{2002PhRvD..66j3511L}, which uses Markov-Chain Monte Carlo
techniques.  We include data from WMAP7 \citep{2010arXiv1001.4635L} and,
separately, ACT \citep{2011ApJ...729...62D} and the dusty star-forming
galaxy-subtracted data from SPT \citep{2010arXiv1012.4788S}.  We fit for 6 basic
cosmological parameters ($\Omega_{\rmn{b}}h^2$, $\Omega_{\rmn{DM}}h^2$,
$n_{\rmn{s}}$, the primordial scalar power spectrum amplitude $A_{\rmn{s}}$, the
Compton depth to re-ionization $\tau$, and the angular parameter characterizing
the sound crossing distance at recombination $\theta$) with the assumption of
spatial flatness. We also include a white noise template for point sources
$C_{\ell,\mathrm{src}}$ with amplitude $A_{\mathrm{src}}$.  The primary
difference between our analysis and the analysis by SPT
\citep{2010arXiv1012.4788S} is that we marginalize over $A_{\mathrm{src}}$,
allowing for arbitrary (positive) values, and ignore the spatial clustering
component of point sources.  We assume a perfect degeneracy
$C_{\ell,\mathrm{kSZ}} \propto C_{\ell,\mathrm{tSZ}}$ for the kinetic SZ (kSZ)
component, so we only need the relative amplitude of
$A_{\mathrm{kSZ}}/A_{\mathrm{tSZ}}$ at a given frequency and use the kSZ
amplitudes from \citet{2010ApJ...725...91B}, where the ratio of kSZ to tSZ at
$\ell=3000$ and 150 GHz is 0.44.  As mentioned in \citet{2010ApJ...725...91B},
these simulations do not fully sample the long wavelength tail of the velocity
power spectrum and do not include any contributions from patchy re-ionization
\citep{2007ApJ...660..933I,2008MNRAS.384..863I}. Hence this kSZ power spectrum
template is a lower limit to the total power.

\begin{figure}
\epsscale{1.2}
\plotone{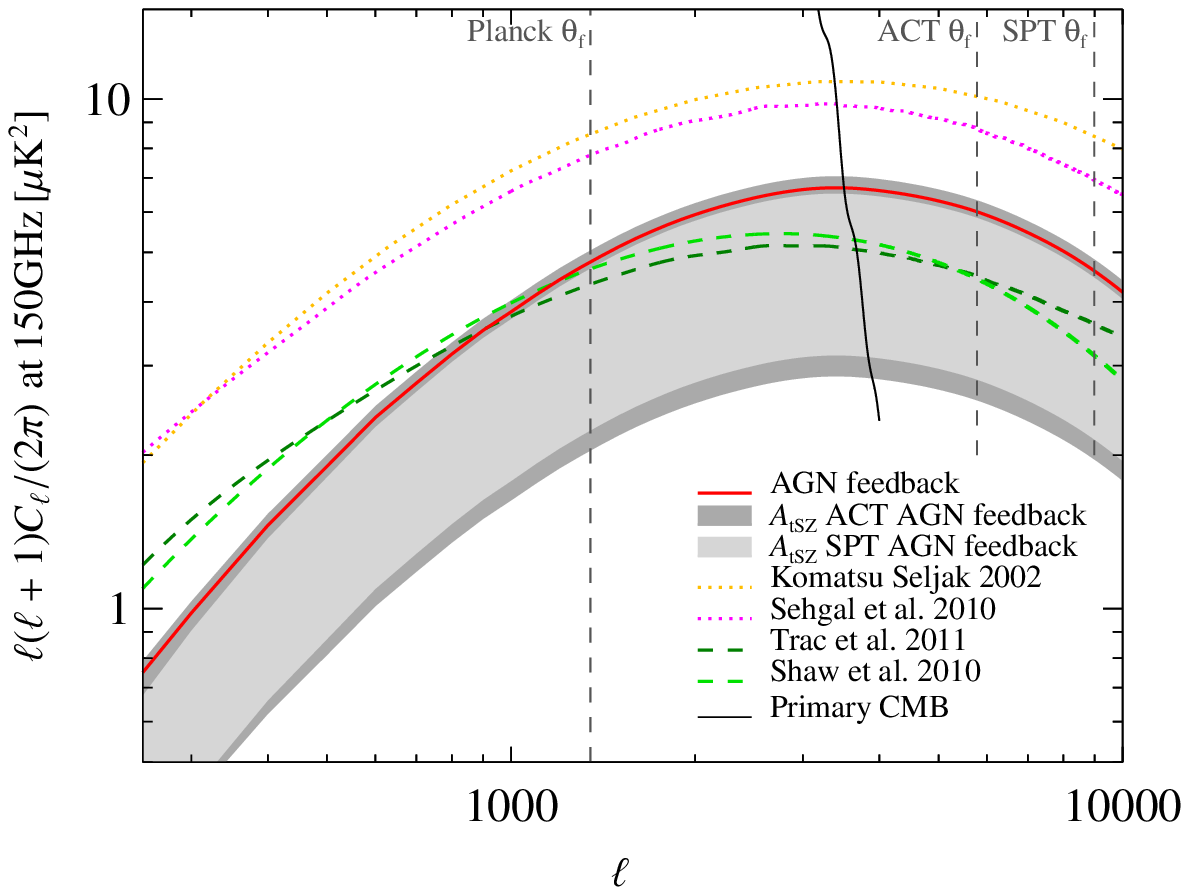}
\caption[The constraints on the amplitude of the thermal Sunyaev-Zel'dovich
power spectrum from current ACT and SPT measurements]{Our 150 GHz tSZ power
  spectrum of our AGN feedback model, rescaled to the
  \citet{2011arXiv1105.3182K} best-fit $\sigma_8$ value of 0.814 (red line) is
  contrasted with the bands indicating the $68$\% range in tSZ amplitude from
  ACT \citep[][dark grey]{2011ApJ...729...62D} and SPT \citep[][light
  grey]{2010arXiv1012.4788S}.  For comparison, we plot several other models for
  the tSZ power spectrum, also shifted to the fiducial $\sigma_8=0.814$.  These
  are \citet{2010ApJ...709..920S} (pink dotted line),
  \citet{2011ApJ...727...94T} (dark green dashed line),
  \citet{2002MNRAS.336.1256K} (orange dotted line), and
  \citet{2010ApJ...725.1452S} (light green dashed line).  We include the
  estimated beam FWHM for ACT, SPT, and Planck.  }
\label{fig:sig8cont}
\end{figure}

In Figure \ref{fig:sig8cont} we illustrate the 68\% allowed confidence
intervals for the tSZ power spectrum, given the shape of our AGN
feedback template, our predicted tSZ-to-kSZ power spectrum ratio, and
the current data from ACT and SPT. We scale our template using the
best-fit $\sigma_8$ value from \citet{2011arXiv1105.3182K} of 0.814
and scale our template (which was calculated at $\sigma_8=0.8$) by
$(0.814/0.8)^8$, about 15\%.  We find that our template is within about the
68\% confidence interval region for both ACT and SPT, after correcting
for our predicted kSZ to tSZ power spectrum ratio of 0.44.  
Note that the semi-analytic and analytic models
without substructure have lower tSZ amplitudes, which would result in
higher values of $\atsz$ and higher $\sigma_8$. 

\section{Discussion and Conclusion}
\label{sec:con}

In this work, we found a global fitting function for cluster thermal pressure
profiles using the simulations presented in \citet{2010ApJ...725...91B}.  We
find that this global fit matches the mean pressure profiles across mass and
redshift generally to an accuracy of better than 10\%.  We have used the profile
fit to reconstruct the thermal Sunyaev-Zel'dovich power spectrum using both
fully analytic and semi-analytic pasted profiles onto cluster position in the
simulations, and find we recover the tSZ power spectrum to $\sim$15\% at
$\ell=3000$ (cf. Figure \ref{fig:clcomp}).  Other analytic and semi-analytic
models for the tSZ effect commonly assume a constant logarithmic slope of the
$P$--$\rho$ relation, $\Gamma$, when solving the equation of hydrostatic
equilibrium.  The assumption is not borne out in our simulations, where both the
thermal $\Gamma$ (which account only for the thermal pressure) and the effective
pressure $\Gamma$ (which includes non-thermal support from bulk flows in
clusters to the pressure) considerably increase in cluster outskirts (cf. Figure
\ref{fig:clcdf}).  Using both the simulations and the global pressure profile,
we examined the contributions to the tSZ spectrum as functions of cluster mass,
redshift, and truncation radius.  We found that the contributions from
substructure and asphericity are most important for the highest mass clusters
($M_{500}\gtrsim 4.2\times 10^{14} M_{\sun}$), but remain significant at the
$10-15$\% level across all mass bins.  We find that half the power of the tSZ
power spectrum at $\ell=3000$ is contributed by clusters with $z >0.6$ and half
the power originates from clusters with $M_{500}<2\times 10^{14}M_{\sun}$.

We have compared our tSZ prediction to results from the Atacama
Cosmology Telescope and the South Pole Telescope.  We found that there
is no statistically significant difference between our model and the
data, after accounting for a simplistic correction from the kinetic SZ
effect.  More complete component separation should be possible with
better frequency coverage \citep{2011arXiv1102.5195M}.  We note that
our analysis differs from that in \citet{2010arXiv1012.4788S} in that
we make no prior assumption about the amplitude of the point source
power spectrum, other than that it is non-negative.

The pressure profile presented in this work is derived from the {\it
  mean} electron pressure in our simulations and, as such, is
appropriate for comparison with individual clusters; we defer the
derivation of a mean profile designed to include the effects of
substructure and asphericity in the power spectrum to a future work
(Battaglia et al., in prep).  This profile will not be expected to
match individual cluster observations, but we hope will allow analytic
calculations of the tSZ power spectrum to an accuracy of significantly
better than 10\%.  With future data sets, such as those expected from
Planck, ACTpol, and SPTpol, it may be possible to constrain not just
the amplitude but the shape of the tSZ spectrum. In this case,
analytic calculations may be usable to constrain not just cosmology
but the important astrophysical processes in clusters with the tSZ
effect.  Doing so through the power spectrum has the advantage that it
is sensitive to lower mass and higher redshift clusters as well as cluster
outskirts in ways that are complementary to other data sets.

\begin{table}
  \label{tab:sig8}
  \caption{Cosmological constraints on $\atsz$ and $\sigma_8$ from ACT
  and SPT using the AGN feedback tSZ power spectrum template}
  \begin{center}
   \leavevmode
   \begin{tabular}{lcc} 
     \hline \hline              
     Data & $\atsz$ & $\sigma_8$ \\
     \hline
     ACT \citep{2011ApJ...729...62D}   & $0.85 \pm 0.36$ & $0.784^{+0.036}_{-0.053}$\\ 
     SPT \citep{2010arXiv1012.4788S}   & $0.69 \pm 0.29$ & $0.764^{+0.035}_{-0.051}$\\        
     \hline
    \end{tabular}
  \end{center}
\end{table}

\acknowledgments

We thank Norm~Murray, Neal~Dalal, Mike~Nolta, Phil~Chang, Hy~Trac,
Diasuke~Nagai, Laurie~Shaw, Doug~Rudd, and Gus~Evrard for their useful
discussions. Research in Canada is supported by NSERC and
CIFAR. Simulations were run on SCINET and CITA's Sunnyvale
high-performance computing clusters. SCINET is funded and supported by
CFI, NSERC, Ontario, ORF-RE and UofT deans.  C.P. gratefully
acknowledges financial support of the Klaus Tschira Foundation.
We also thank KITP for their hospitality during the
galaxy cluster workshop. KITP is supported by National Science
Foundation under Grant No. NSF PHY05-51164

\bibliography{bibtex/nab}
\bibliographystyle{apj}


\begin{appendix}

\section{Comparing the Cluster Mass Catalog to the Mass function}

\begin{figure}
\epsscale{1.2}
\plotone{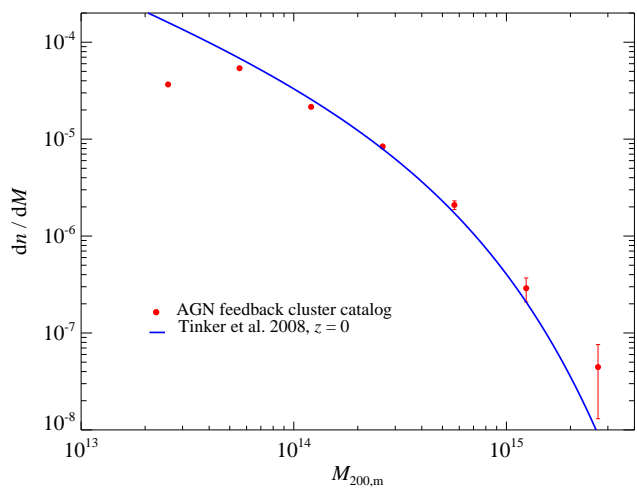}
\caption[Comparison of the clusters catalog from AGN feedback
simulations to the mass function from \citet{2008ApJ...688..709T}]{We
compare the mass function, $\dd n\,/\,\dd M$, for the cluster catalog from AGN feedback
simulations to the mass function from \citet{2008ApJ...688..709T}. The
differences at high masses indicates that in the 10 independent
simulations we happen to have more high-mass clusters than is
expected on average (though with only 6 with $M_{500} > 7.1 \times
10^{14} M_{\sun}$ this is consistent with shot noise). At low masses,
our catalog is incomplete due to our FOF halo finding (see text).
}
\label{fig:mf}
\end{figure}

In this appendix we compare the mass function from our simulations
with that of \citet{2008ApJ...688..709T}.  Our cluster mass catalogs were made
with spherical overdensity mass with respect to the critical density
and the mass function is with respect to the mean matter density. So,
we converted the $M_{200}$ from the simulations to $M_{200,\rmn{m}}$
assuming the mass profile is dominated by dark matter and use the
concentration-mass relations from \citet{2008MNRAS.390L..64D}. We show
in Figure \ref{fig:mf} that there is a clear deficit of low mass 
clusters due to the chosen linking length of 0.2 in our FOF finder.
At this length, is is well known that neighboring clusters are
sometimes artificially merged together
\citep[e.g.,][]{1985ApJ...292..371D}.  We also instituted a firm lower
limit mass cutoff in the initial FOF catalogs of $M_{\rm{FOF}}>1.4 \times 10^{13}
M_{\sun}$, and so our mass function is also expected to be incomplete
near that mass.

There is a clear excess of high-mass clusters in our simulations, but
it is consistent with shot noise (we only have 6 clusters with
$M_{500}>7.1 \times 10^{14} M_{\sun}$).  We now estimate the excess
power in our full simulation power spectrum due to this upwards
fluctuation in the highest mass bin.  Where the cluster catalogs are
complete, we expect that over an enormous number of simulations, the
paste profile and analytic calculation of the tSZ power spectrum would
converge, and indeed see the agreement is excellent between the two in
the right panel of Figure \ref{fig:clmcut} for all but the lowest (due
to catalog incompleteness) and highest (due to shot noise) mass bins.
We therefore adopt the ratio of the the pasted profile spectrum to the
analytic spectrum as a quantitative estimate of the
over-representation of high mass clusters in our finite number of
realizations.  At $\ell=3000$ this ratio is 2.0 for clusters with
$M_{500}>7.1 \times 10^{14} M_{\sun}$, though as can be seen in Figure
\ref{fig:clmcut} the specific value is insensitive to the reference
$\ell$.  Since the high-mass contribution to the tSZ spectrum from the
full simulation projections is $0.67 \mu \rm{K}^2$ at $\ell=3000$, in
the limit of an infinite number of simulations, we would expect the
average contribution from clusters with $M_{500} > 7.1 \times 10^{14}
M_{\sun}$ to be $0.34 \mu \rm{K}^2$ lower.  The total power spectrum
at $\ell=3000$ is $5.78~\mu \rm{K}^2$, so this shot noise correction
amounts to just less than a 6\% shift in the total power spectrum.

\end{appendix}

\end{document}